\documentclass[a4paper,preprintnumbers,showpacs,onecolumn,superscriptaddress,nofootinbib,amsmath,amssymb,notitlepage]{revtex4-2}

\usepackage{mathtools,leftindex,tensor,mhchem}
\usepackage{float}
\usepackage{booktabs}
\usepackage{subfigure}
\usepackage{enumitem}
\usepackage{times}
\usepackage{dcolumn}
\usepackage{mathrsfs}
\usepackage{comment}
\usepackage{multirow}
\usepackage{hyperref}
\usepackage{amsmath,accents}
\usepackage{mathtools}
\usepackage{bbm}
\usepackage{bm}
\usepackage{amsfonts}
\usepackage{amssymb}
\usepackage{mathrsfs}
\usepackage[scr=rsfs,cal=boondox]{mathalfa}
\usepackage{wasysym}
\usepackage{graphicx}
\usepackage{filecontents}
\usepackage[dvipsnames]{xcolor}
\usepackage{bbold}
\usepackage[scr=rsfs,cal=boondox]{mathalfa}

\usepackage{stackengine}
\stackMath
\newcommand\tsup[2][2]{%
 \def\useanchorwidth{T}%
  \ifnum#1>1%
    \stackon[-1.3ex]{\tsup[\numexpr#1-1\relax]{#2}}{\mathchar"307E}%
  \else%
    \stackon[-1ex]{#2}{\mathchar"307E}%
  \fi%
}
\hypersetup{
    bookmarks=true,         
    pdftoolbar=true,        
    pdfmenubar=true,        
    pdffitwindow=true,     
    pdfstartview={FitH},     
    pdftitle={My title},     
    pdfauthor={author},      
    pdfsubject={Subject},    
    pdfcreator={Creator},    
    pdfproducer={Producer},  
    pdfkeywords={keyword1} 
    pdfnewwindow=true,       
    colorlinks=true ,        
    linkcolor=Blue  ,        
    citecolor=Blue,      
    filecolor=Blue,       
    urlcolor=Blue            
}

\newcommand{\oalpha}[1]{\accentset{\circ}{\alpha}}
\newcommand{\obf}[1]{\accentset{\circ}{\mathbf{f}}}
\newcommand{\boR}[1]{\accentset{\circ}{\mathbf{R}}}
\newcommand{\obF}[1]{\accentset{\circ}{\mathbf{F}}}
\newcommand{\obPi}[1]{\accentset{\circ}{\mathbf{\Pi}}}

\usepackage{scalerel}
\usepackage{tikz}
\usetikzlibrary{svg.path}

\definecolor{orcidlogocol}{HTML}{A6CE39}
\tikzset{
  orcidlogo/.pic={
    \fill[orcidlogocol] svg{M256,128c0,70.7-57.3,128-128,128C57.3,256,0,198.7,0,128C0,57.3,57.3,0,128,0C198.7,0,256,57.3,256,128z};
    \fill[white] svg{M86.3,186.2H70.9V79.1h15.4v48.4V186.2z}
                 svg{M108.9,79.1h41.6c39.6,0,57,28.3,57,53.6c0,27.5-21.5,53.6-56.8,53.6h-41.8V79.1z M124.3,172.4h24.5c34.9,0,42.9-26.5,42.9-39.7c0-21.5-13.7-39.7-43.7-39.7h-23.7V172.4z}
                 svg{M88.7,56.8c0,5.5-4.5,10.1-10.1,10.1c-5.6,0-10.1-4.6-10.1-10.1c0-5.6,4.5-10.1,10.1-10.1C84.2,46.7,88.7,51.3,88.7,56.8z};
  }
}

\newcommand\orcidicon[1]{\href{https://orcid.org/#1}{\mbox{\scalerel*{
\begin{tikzpicture}[yscale=-1,transform shape]
\pic{orcidlogo};
\end{tikzpicture}
}{|}}}}

\begin{document} \sloppy
\title{
Shadows aspect of rotating black holes in the Einstein-\textsc{AdS} \textsc{SU}(\textnormal{\textit{N}})-nonlinear sigma model}

\author{Mohsen Fathi\orcidicon{0000-0002-1602-0722}}
\email{ mohsen.fathi@ucentral.cl}
\affiliation{Centro de Investigaci\'{o}n en Ciencias del Espacio y F\'{i}sica Te\'{o}rica, Universidad Central de Chile, La Serena 1710164, Chile}

\author{Yassine Sekhmani\orcidicon{0000-0001-7448-4579}}
\email{sekhmaniyassine@gmail.com}
\affiliation{Center for Theoretical Physics, Khazar University, 41 Mehseti Street, Baku, AZ1096, Azerbaijan.}
\affiliation{Centre for Research Impact \& Outcome, Chitkara University Institute of Engineering and Technology, Chitkara University, Rajpura, 140401, Punjab, India.}

\begin{abstract}

Recent observations of the supermassive black holes M87* and Sgr A* by the Event Horizon Telescope (EHT) have opened new avenues for testing gravity theories through black hole shadow observables. These observations offer a means to distinguish between general relativity and modified gravity theories while providing insights into the astrophysical properties of the observed black holes. In this work, we investigate photon orbits and shadow characteristics of rotating black holes within the Einstein-$\mathrm{SU}(N)$ nonlinear sigma model. We first analyze the static, spherically symmetric black hole solution, focusing on its asymptotically Anti-de Sitter behavior and causal structure. The Modified Newman-Janis Algorithm is then applied to obtain the rotating counterpart, followed by an examination of its geometric properties, ergoregions, and causal structure. Utilizing the Lagrangian formalism, we derive the equations of motion for photons and study the resulting black hole shadow on the celestial plane. We explore the dependence of the shadow's size and shape on the black hole parameters $K$ and $N$, imposing constraints using EHT observational data for M87* and Sgr A*. Finally, we analyze the black hole's evaporation rate under different scenarios.

\end{abstract}

\maketitle

\section{Introduction}

 Black holes (BHs), first predicted by general relativity (GR), are among the most fascinating objects in the universe and serve as vital testing grounds for various theories of gravity. They are particularly important in extreme conditions where both GR and quantum effects are relevant. A defining feature of BHs is their event horizon, which acts as a boundary from which no information, including light, can escape \citep{Schwarzschild:1916uq}. The formation of BHs is a natural consequence of GR \citep{Penrose:1964wq}, and they remain central to efforts aimed at reconciling GR with quantum mechanics \citep{Hawking:1976ra}. The no-hair theorem, a foundational principle in BH physics under GR, posits that any stationary, asymptotically flat, axially symmetric BH solution to the Einstein field equations \citep{CarterPRD1968} can be fully described by three parameters: mass $M$, angular momentum $j$, and electric charge $Q$ \citep{Israel:1967wq, Israel:1967za, Carter:1971zc, Hawking:1971vc, Robinson:1975bv}. These parameters are encapsulated in the Kerr-Newman metric \citep{Newman:1965my}, which generalizes the Kerr solution \cite{Kerr:1963ud} (a solution for a rotating, electrically neutral BH). 
The no-hair theorem is based on the uniqueness theorem of GR, which states that the Kerr-Newman metric is the sole solution to the Einstein-Maxwell equations that satisfy the conditions of stationarity and axial symmetry. 
For practical applications, BHs are often considered electrically neutral, as any residual charge is quickly neutralized through interactions with surrounding matter \citep{Israel:1967wq, Israel:1967za, Carter:1971zc, Carter:1999mrq, Hawking:1971vc}. 

There has been some debate about the mathematical foundation of the no-hair theorem, especially concerning the assumption of analyticity \citep{Chrusciel:2012jk}. The assumption that smooth, analytic metrics are accurate representations of BHs is crucial to the theorem, but it may not be universally applicable in all theoretical contexts. Modifications to GR, such as those found in metric affine gravity theories, allow for BHs that deviate from the Kerr solution. Testing the no-hair theorem thus involves exploring possible deviations from Kerr metrics, particularly in scenarios proposed by modified theories of gravity (MTG).
Consequently, the Kerr metric is commonly used to model observed BHs. Disproving the existence of non-Kerr BHs requires both observational and theoretical efforts \citep{Ryan:1995wh, Will:2005va}.

The study of BHs has seen significant advancements, especially following the groundbreaking detection of gravitational waves by LIGO in 2015. Additionally, the EHT has utilized the Very Long Baseline Interferometry (VLBI) technique to obtain the high angular resolution needed to image supermassive BHs. VLBI at a wavelength of 1.3 mm (230 GHz) with Earth-diameter-scale baselines is essential for resolving the shadows of M87* and Sgr A*, the two supermassive BHs with the largest apparent angular sizes \citep{Johannsen:2012vz}. The EHT's development has led to the first images of BH shadows, from M87* \citep{EventHorizonTelescope:2019dse, EventHorizonTelescope:2019pgp, EventHorizonTelescope:2019ggy,EventHorizonTelescope:2019jan, EventHorizonTelescope:2019ths, EventHorizonTelescope:2019uob} and Sgr A* \citep{EventHorizonTelescope:2022exc,EventHorizonTelescope:2022urf,EventHorizonTelescope:2022apq,EventHorizonTelescope:2022wok,EventHorizonTelescope:2022wkp, EventHorizonTelescope:2022xqj}, significantly enhancing our ability to test theories of gravity { \citep{CapozzielloJCAP2023,Capozziello2024,HuangJCAP2024,macedo_optical_2024} and demonstrating the alignment of the observed shadows with predictions for a Kerr BH in GR. 
In turn, investigating photon motions inside the Kerr BH is very pertinent to current findings. A specific category of photon orbits with fixed radii, known as spherical photon orbits, delineates the shadow silhouette and has astrophysical significance inside the photon orbits around a rotating BH spacetime \citep{WilkinsPRD1972}. The BH shadow refers to the dark region seen against the bright background of the accretion disk and surrounding space \citep{PERLICK20221,ZareJCAP2024,MengPRD2023,FENG2025103075,MengPRD2022,UNIYAL2023101178,ZAREPLB2024}. Photons that come extremely close to the BH, near the photon sphere, create this shadow, which is surrounded by a bright ring of light, the photon ring \citep{Synge:1966okc,Bardeen:1973tla,Luminet:1979nyg,Cunningham:1973tf}, as seen on the observer's sky, with its outline marked by gravitationally lensed photons \citep{Johannsen:2010ru}. 
The study of BH shadows has become an essential tool for understanding the near-horizon geometry, and much research has focused on analyzing these shadows in the context of both GR \citep{Falcke:1999pj,Shen:2005cw,Hioki:2009na,Yumoto:2012kz,Atamurotov:2013sca,Abdujabbarov:2015xqa,Cunha:2018acu,Kumar:2018ple,mishra_understanding_2019,Afrin:2021ggx,khodadi_no-hair_2021,khodadi_shadow_2022,Chen:2023wzv,Li:2024abk} and MTG \citep{Amarilla:2010zq,Amarilla:2011fx,Amarilla:2013sj,Papnoi:2014aaa,Amir:2017slq,Singh:2017vfr,Mizuno:2018lxz,Allahyari:2019jqz,Vagnozzi:2019apd,Kumar:2020hgm,Kumar:2020owy,Ghosh:2020spb,Guo:2020zmf,khodadi_black_2020,Afrin:2021wlj,Vagnozzi:2022moj,khodadi_probing_2022,olmo_shadows_2023,asukula_spherically_2024,khodadi_event_2024}, including in quantum gravity inspired frameworks \citep{Liu:2020ola,Brahma:2020eos,KumarWalia:2022ddq,Islam:2022wck,Afrin:2022ztr,Yang:2022btw,battista_quantum_2024,wang_dynamical_2025}. The study of BH shadows has also extended to higher-dimensional spacetimes within modified gravity theories, where extra dimensions are introduced \citep{Papnoi:2014aaa,Singh:2017vfr,Amir:2017slq,Vagnozzi:2019apd,banerjee_silhouette_2020,banerjee_hunting_2022}. BH shadows thus provide a powerful tool for strong-field gravitational tests and, ultimately, for probing the validity of the no-hair theorem \citep{Johannsen:2010ru,Baker:2014zba,Glampedakis:2023eek,Afrin:2021imp,Afrin:2021wlj,Afrin:2022ztr,Islam:2022wck}.

At the same time, the nonlinear sigma model (NLSM) has become a crucial effective field theory, both in theoretical and phenomenological terms. It has widespread applications in quantum field theory, statistical mechanics, and string theory \citep{manton2004topological}. One of the key uses of the NLSM is describing the low-energy dynamics of pions \citep{nair2006quantum}, typically in the context of the internal symmetry group $ \mathrm{SU(2)} $, which represents the two-flavor case. While the predictions made by the NLSM align well with experimental results, solving the field equations derived from the model can be complex, as these equations are generally nonlinear and coupled, involving $(N^2-1)$ equations where $N$ is the flavor number encoded in the $ \mathrm{\mathrm{SU}(N)} $ group. Consequently, many solutions have been constructed numerically. Some important results in this field are presented in Refs. \citep{Heusler:1991xx,Luckock:1986tr,Glendenning:1988qy,Nelmes:2011zz}. When the NLSM is coupled with GR or Maxwell theory to describe more intricate physical systems, the resulting field equations become even more complex. Nonetheless, in Refs. \citep{8,9}, ansatzs have been proposed that go beyond spherical symmetry (the generalized hedgehog ansatz), which have been used to find exact solutions not only in the NLSM, but also in the Skyrme model, the generalized Skyrme model, and the Yang-Mills-Higgs theory. These solutions describe a variety of objects, including boson stars \citep{10}, BHs \citep{9,11,12}, black strings \citep{14}, gravitating solitons \citep{15,16}, topological solitons at finite volume \citep{17}, and crystalline structures of topological solitons \citep{18,19}.

In this paper, motivated by the study of BHs in MTG, we consider a static spherically symmetric BH solution in an asymptotically Anti-de Sitter (AdS) spacetime, derived from the Einstein-$\mathrm{SU}(N)$-NLSM, as proposed in Ref. \citep{Henriquez-Baez:2022ubu}. We then apply a modified version of the Newman-Janis algorithm (NJA) \cite{NewmanJanis:1965} to obtain the rotating counterpart. The modified NJA (MNJA), as discussed in Refs. \cite{Azreg:2014, Azreg:2014_1} which proves particularly effective when applied to theories extending beyond GR, offering enhanced accuracy and stability in these frameworks. The primary objective of this study is to perform both analytical and numerical investigations of photon orbits and the shadow of the rotating BH spacetime, with the goal of constraining its parameters based on the observational data from the EHT.

The structure of the paper is as follows: In Sect. \ref{sec:Model}, 
we explore the non-rotating BHs in the Einstein-$\mathrm{SU}(N)$-NLSM and investigate the influence of the model parameters on the horizon structure. The causal structure of this BH is then examined. We proceed by applying the MNJA to derive the rotating counterpart of the BH spacetime, referred to as the rotating Einstein-AdS-$\mathrm{SU}(N)$-NLSM BH (RASN-BH), and analyze its causal structure and ergoregions. 
In Sect. \ref{sec:photons}, we utilize the standard Lagrangian formalism to derive the equations of motion for photons traveling through the exterior geometry of the RASN-BH. In this section, we parametrize the BH shadow on a two-dimensional celestial plane and explore how the shadow is influenced by the choice of BH parameters. 
Sect. \ref{sec:observables} presents the main observables associated with deformed shadows, and each of these is discussed individually. 
These observables are then used to constrain the BH parameters based on the EHT observational data for M87* and Sgr A*. In Sect. \ref{sec:energyEmission}, we discuss the dependence of the BH evaporation rate on the spacetime parameters. Finally, we conclude in Sect. \ref{sec:conclusions}.

Throughout the paper, we adopt geometric units with $G = c = 1$, and use the $(-,+,+,+)$ sign convention. Primed symbols denote derivatives with respect to the radial coordinate.

\section{The Einstein-$\mathrm{SU}(N)$-NLSM and its A\lowercase{d}S BH solution}\label{sec:Model} 

An analysis of BH solutions within the framework of the Einstein-$\mathrm{SU}(N)$-NLSM is crucial for evaluating the physical viability of the model. The Einstein-$\mathrm{SU}(N)$-NLSM is characterized by the action \cite{Henriquez-Baez:2022ubu}
\begin{equation}\label{action}
    S_{g,U} = \int \mathrm{d}^4x \sqrt{-g} \left( \frac{1}{2\kappa} (R - 2\Lambda) + \frac{K}{4} \mathrm{Tr}[L^\mu L_\mu] \right),
\end{equation}
where $R$ denotes the Ricci scalar, $\Lambda$ represents the cosmological constant, and $L_\mu$ corresponds to the components of the Maurer-Cartan form, expressed as
\begin{equation}
    L_\mu = U^{-1} \partial_\mu U = L_\mu^i t_i.
\end{equation}
Here, $U(x) \in \mathrm{SU}(N)$, with $N$ denoting the number of flavors embedded within the $\mathrm{SU}(N)$ Lie group \cite{Bertini:2005rc,Cacciatori:2012qi,Tilma:2004kp}, and $t_i$ being the generators of the $\mathrm{SU}(N)$ Lie algebra, where $i = 1, \dots, (N^2 - 1)$. In this context, $\kappa$ is the gravitational constant, while $K$ is a positive coupling constant determined through experimental observations. 

The Einstein-$\mathrm{SU}(N)$-NLSM framework gives rise to deformed Schwarzschild-AdS BHs \cite{Henriquez-Baez:2022ubu}, describing an asymptotically AdS, static, and spherically symmetric BH spacetime. The corresponding line element is given by
\begin{equation}\label{metric}
    \mathrm{d}s^2 = -f(r) \mathrm{d}t^2 + \frac{1}{f(r)} \mathrm{d}r^2 + r^2 \mathrm{d}\theta^2 + r^2 \sin^2\theta \mathrm{d}\phi^2,
\end{equation}
where the metric function $f(r)$ is defined as
\begin{equation}\label{e2}
   f(r) = 1 - \frac{2M}{r} - \frac{\Lambda}{3}r^2 - K\,\kappa\, a_N.
\end{equation}
Here, $M$ represents the BH mass, while the positive quantity $a_N$ is given by
\begin{equation} \label{aN}
 a_N = \frac{N(N^2-1)}{6}.
\end{equation}
{One important property of the lapse function \eqref{e2} is the presence of a deficit angle, which is explicitly manifested in the term $K\kappa a_N$. This characteristic is commonly observed in black hole spacetimes associated with a cloud of strings (see, for instance, Refs. \citep{letelier_clouds_1979,toledo_black_2018,morais_graca_cloud_2018}). It has been demonstrated that the deficit angle can influence the motion of both massive and massless test particles by appearing in the equations of motion (see, e.g., Refs. \citep{batool_null_2017,mustafa_radial_2021,fathi_study_2022}).
}

The notable characteristics of this class of BHs are twofold: first, they are supported by pionic matter, and second, they generalize the $\mathrm{SU}(N)$ BH analyzed in \cite{Canfora:2013osa} (excluding the Skyrme term) and \cite{Gibbons:1990um}. Notably, this BH solution exhibits an asymptotic structure corresponding to the AdS version of the Barriola-Vilenkin metric \cite{Rhie:1990kc} and reduces to the Schwarzschild-AdS spacetime in the limit $K = 0$.

It is evident that the metric given by Eq. \eqref{e2} has a coordinate singularity at $f(r) = 0$, leading to only one real positive root denoted by $r_{+}$, which corresponds to the radius of the BH event horizon, give as
\begin{equation}\label{root}
 r_+ = \frac{\kappa K a_N \Lambda - {\cal P}^2 - \Lambda}{\Lambda {\cal P}},
\end{equation}
where ${\cal P}^3 = \sqrt{\Lambda^{3} \left[ \left(\kappa K a_N - 1 \right)^3 + 9 \Lambda M^2 \right]} + 3 \Lambda^{2} M$. 
The geometry associated with the BH spacetime in the Einstein-$\mathrm{SU}(N)$-NLSM, has been then shown in Fig. \ref{fig11}, where the behavior of the metric function $f(r)$ under variations of different parameter values while keeping the others fixed. 
\begin{figure}[h]
	\centering
	\includegraphics[width=7.5cm,height=6cm]{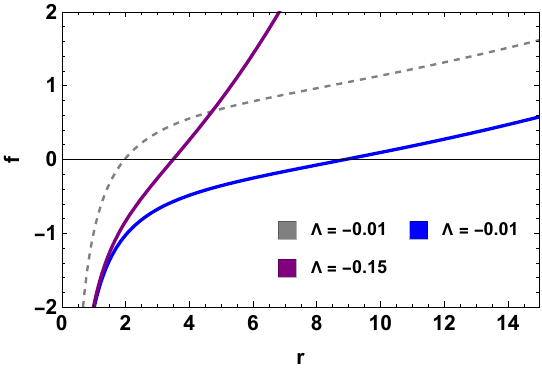} \qquad 
	\includegraphics[width=7.5cm,height=6cm]{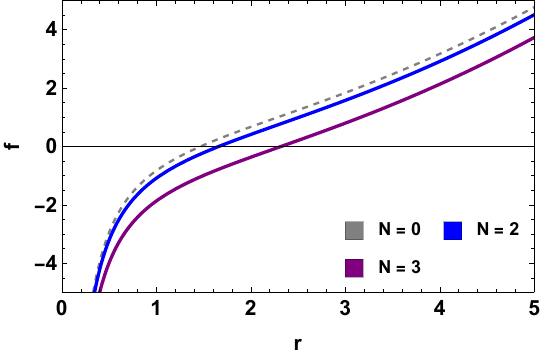} 
        \includegraphics[width=7.5cm,height=6cm]{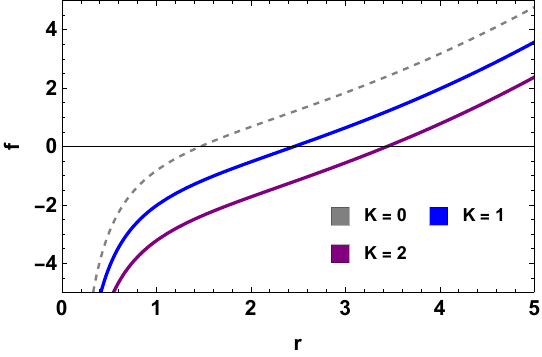}  \qquad
        \includegraphics[width=7.5cm,height=6cm]{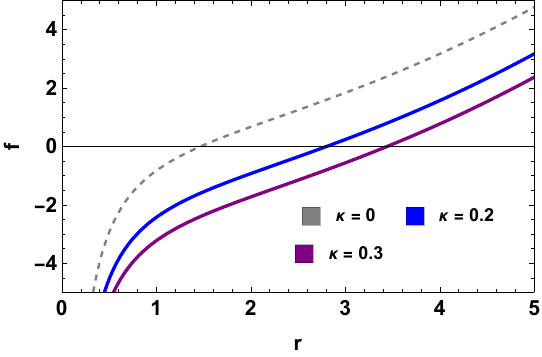}  
	\caption{The plot illustrates the behavior of the metric lapse function $f(r)$ for different values of the BH parameters $\Lambda$, $N$, $K$, and $\kappa$. The length unit along both axes is set by the BH mass $M$.}
	\label{fig11}
\end{figure}
It is evident that $f(r)$ exhibits significant changes for larger values of the model parameters, both in the vicinity of the BH and at larger distances.

To further investigate the properties of this BH solution, modeled by the metric function, we analyze curvature-based tools in the subsequent section.

The Kretschmann scalar is explicitly given by
\begin{equation}\label{Kr}
\mathcal{K} = R^{\mu\nu\alpha\beta} R_{\mu\nu\alpha\beta} = \frac{\kappa^2 K^2 \left(N^3 - N\right)^2 r^2 + 24 \kappa K M N \left(N^2 - 1\right) r + 4 \kappa K \Lambda N \left(N^2 - 1\right) r^4 + 432 M^2 + 24 \Lambda^2 r^6}{9 r^6}.
\end{equation}
A straightforward analysis of the Kretschmann scalar with $K=0$ and $\Lambda=0$ shows that it reduces to ${48M^2}/{r^6}$, which corresponds to the Kretschmann scalar of the Schwarzschild BH. This function, as a function of the radius $r$, is plotted for different values of the flavor number $N$ in Fig. \ref{Kretsch}.
\begin{figure}[h!]
	\centering
	\begin{minipage}[t]{0.5\textwidth}
		\centering
		\includegraphics[width=\textwidth, keepaspectratio]{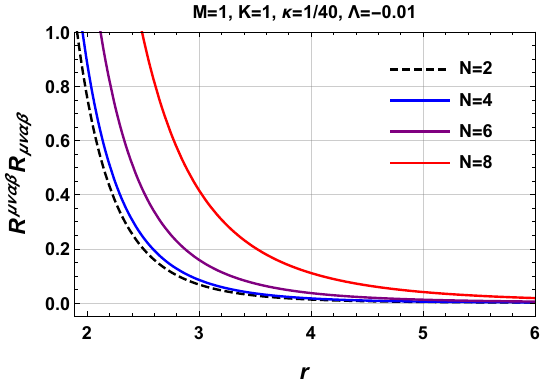}
	\end{minipage}
	\caption{The Kretschmann scalar as a function of the radial coordinate $r$ for various values of the flavor number $N$. The length unit along both axes is set by the BH mass $M$.}
	\label{Kretsch}
\end{figure}
It is important to note that the plot illustrates the asymptotic behavior of the metric, which closely resembles the Schwarzschild spacetime. Consequently, the $\mathcal{K}$-singularity occurs at $r=0$, similar to the Schwarzschild BH.

\subsection{The RASN-BH solution}

The rotating counterpart of the spacetime given in Eqs. \eqref{metric} and \eqref{e2} is expressed as (see Appendix \ref{app:A})
\begin{eqnarray}\label{e1}
	\mathrm{d}s^2 =  \frac{\Delta_\theta \sin^2\theta}{{\Sigma^{2}}\rho^2} \Bigl( a \, \mathrm{d}t - \left(r^2 + a^2\right) \, \mathrm{d}\phi \Bigr)^2 + \frac{\rho^2}{\Delta_r} \, \mathrm{d}r^2 + \frac{\rho^2}{\Delta_\theta} \, \mathrm{d}\theta^2-\frac{\Delta_r}{{\Sigma^{2}}\rho^2} \Bigl( \mathrm{d}t - a \sin^2\theta \, \mathrm{d}\phi \Bigr)^2,
\end{eqnarray}
in which the coordinates $t$ and $r$ span all of $\mathbb{R}$ while $\theta \in [0, \pi]$ and $\phi \in [0, 2\pi]$, and we have
\begin{subequations}
\begin{align}
	& \Delta_r =\left(r^2+a^2\right)\left(1-\frac{\Lambda}{3}r^2\right)-2 M r-\frac{1}{6} \kappa  K N \left(N^2-1\right) r^2,\label{RASN metric}\\
	& \rho^2 = r^2+a^2 \cos^{2}\theta, \\
   & \Delta_\theta = 1+ \frac{a^2}{3}\Lambda \cos^2\theta,\\
& \Sigma = 1+\frac{\Lambda}{3}a^2.
    \end{align}
    \label{eq:ingredients}
\end{subequations}
Here, $a$ denotes the BH spin parameter, while {The parameters $K$, $N$, and $\kappa$ characterize the deviations of the RASN-BH from the standard Kerr-AdS BH spacetime and are directly linked to the deficit angle present in the seed metric function in Eq. \eqref{e2}. As analyzed in this paper, each of these parameters has distinct effects on the spacetime structure, leading to modifications in the shape of the black hole shadow. These effects will be discussed in detail in the following sections.} 

The coefficients of the metric \eqref{metric} are independent of the time coordinate $t$ and the azimuthal angle $\phi$, implying that $\partial_t$ and $\partial_\phi$ are Killing vector fields. Consequently, any linear combination of these two Killing vector fields is also a Killing vector \cite{Visser:2007fj}. Specifically, the Kerr BH is recovered in the special case where $K = \Lambda = 0$, and the static metric \eqref{metric} with the lapse function \eqref{e2} is obtained when $a = 0$. In Fig. \ref{deltaa}, the radial profile of the $\Delta_r$ function has been plotted, showing its sensitivity to changes in the BH parameters.
\begin{figure}[h]
	\centering
	\includegraphics[width=7.5cm,height=6cm]{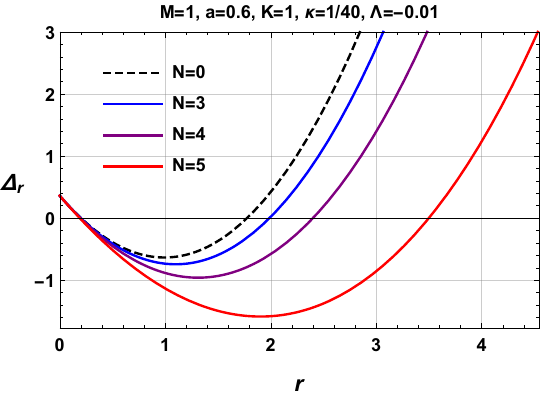} (a)\qquad
	\includegraphics[width=7.5cm,height=6cm]{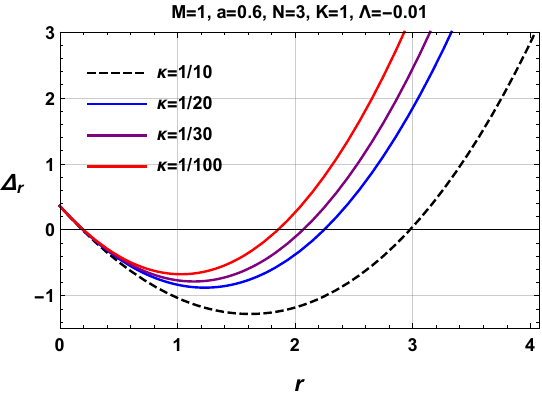} (b)
        \includegraphics[width=7.5cm,height=6cm]{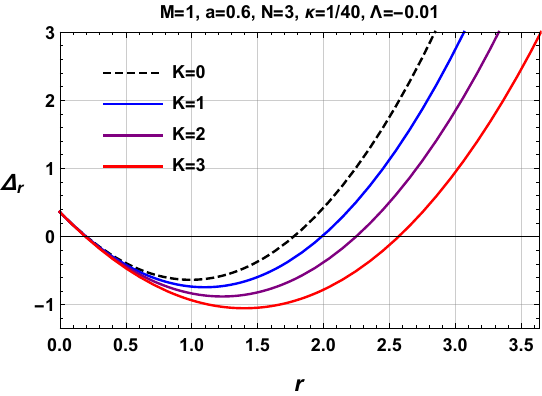} (c)\qquad
	\includegraphics[width=7.5cm,height=6cm]{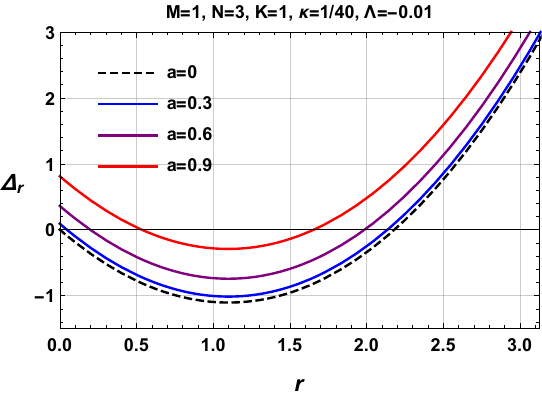} (d)
	\caption{Plots showing the radial profile of $\Delta_r$ (horizons), for the RASN-BH with (a) varying  $N$, (b) varying $\kappa$, (c) varying $\lambda$, and (d) varying $a$. The unit of length is chosen to be the BH mass, $M$.}
	\label{deltaa}
\end{figure}

To further investigate the behavior of the BH system, we show the moduli space corresponding to the parameter spaces $(N, a/M)$ and $(K/M, a/M)$ in Fig. \ref{ps}. In the conventional BH framework, the colored region represents the parameter space with two horizons. Along the black dashed line, for all parameter values $(N_E, (a/M)_E)$ and $((K/M)_E, (a/M)_E)$, the radii of the horizons satisfy $r_+ = r_-$, indicating the extremal BH case. In contrast, the white region represents the space where no physical horizon exists, corresponding to a naked singularity.
\begin{figure}[h]
	\centering
	\includegraphics[width=6.5cm]{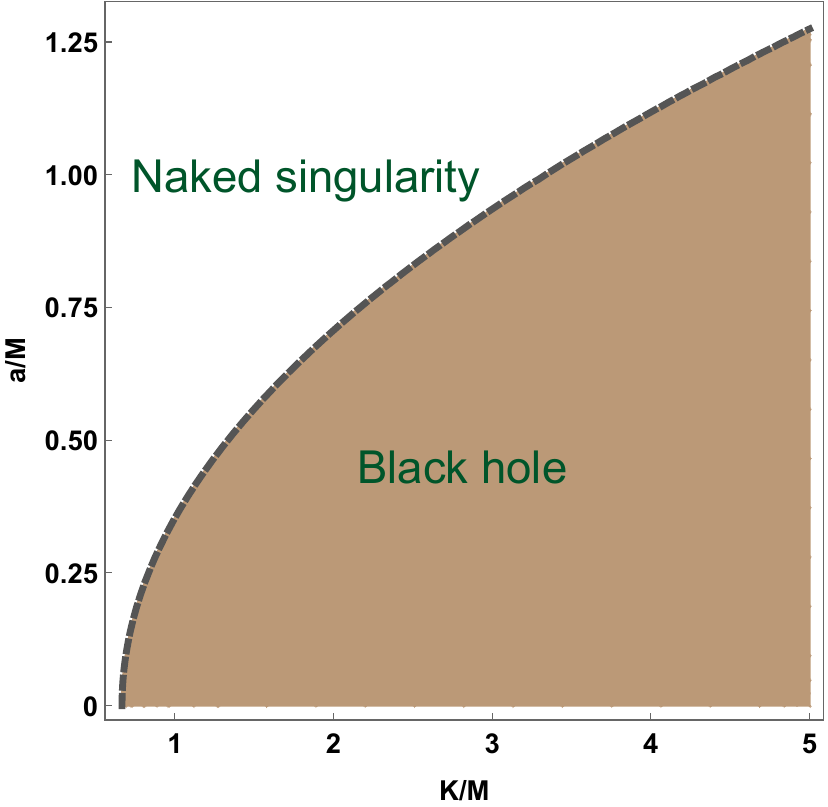} (a)\quad
	\includegraphics[width=6.5cm]{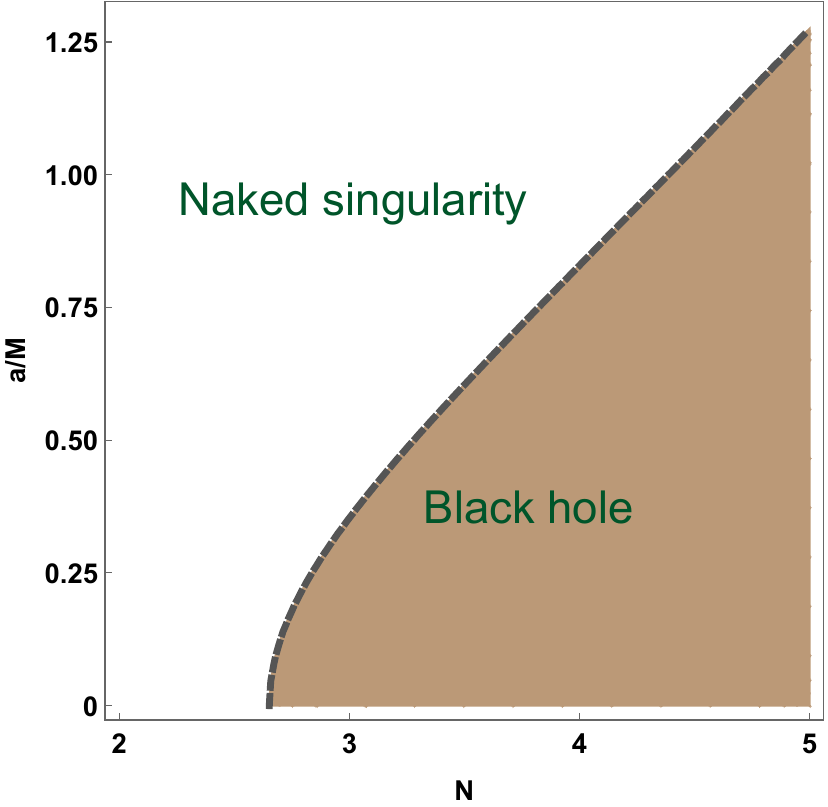} (b)
	\caption{The parameter spaces $(K/M,a)$ and $(N,a/M)$ for the RASN-BH.}
	\label{ps}
\end{figure}

The horizons of the metric \eqref{e1} correspond to the null hypersurfaces defined by the BH's null geodesics, which are determined by solving the equation $g^{\gamma\xi}\partial_\gamma r \partial_\xi r = g^{rr} = \Delta_r = 0$ with $\rho^2 \neq 0$. This equation can yield up to two distinct real positive roots, $r_{\pm}$, or possibly none, depending on the values of the parameters in the rotating case. The root $r_+$ corresponds to the outer (event) horizon, while $r_-$ corresponds to the inner (Cauchy) horizon (see Fig. \ref{ps}). Specifically, when $K = 0$ and $\Lambda = 0$, the solution simplifies to
\begin{equation}
    r_{\pm}=M\pm\sqrt{M^2-a^2},
\end{equation}
where $r_\pm$ represent the horizons of the Kerr BH, provided that $a \leq M$. The RASN metric (\ref{RASN metric}) can model a non-extremal BH when $r_+ > r_-$, while for $r_+ = r_-$, it describes an extremal BH. The behavior of these horizons as a function of changes in the spin parameter is illustrated in Fig. \ref{fig3}.
\begin{figure}[h!]
	\centering
	\includegraphics[width=7.5cm,height=6cm]{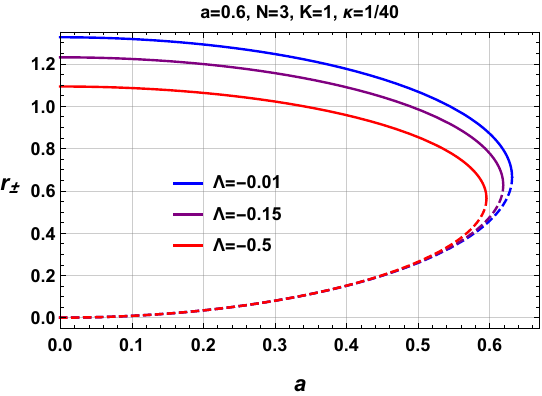} \qquad 
	\includegraphics[width=7.5cm,height=6cm]{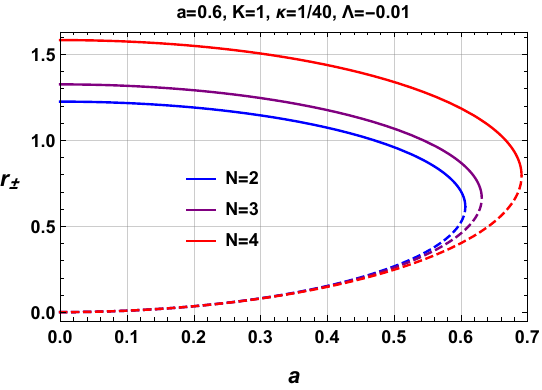} 
         \includegraphics[width=7.5cm,height=6cm]{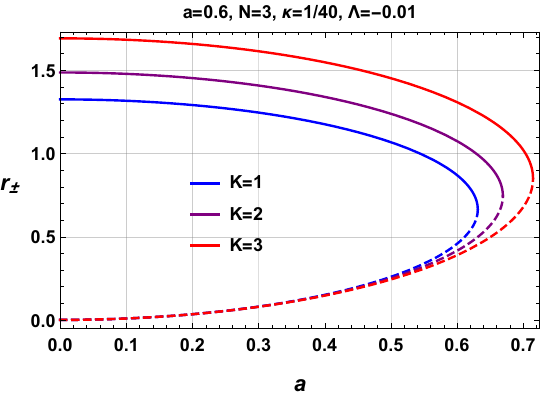}  \qquad
         \includegraphics[width=7.5cm,height=6cm]{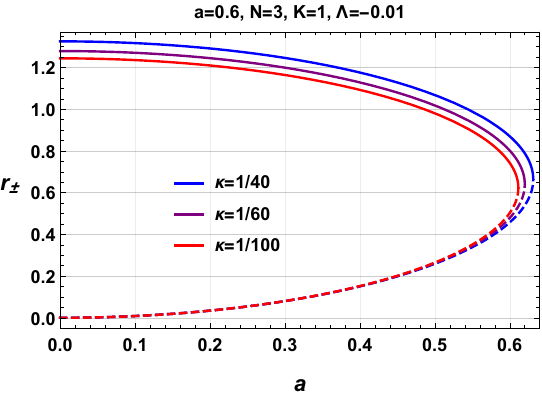}  
	\caption{Event horizons (solid curves) and Cauchy horizons (dashed curves) of the RASN-BH for different values of the BH parameters. The unit of length is chosen to be the BH mass, $M$.}
	\label{fig3}
\end{figure}

At the static limit surface (SLS), the asymptotic time-translational Killing vector $\chi^{i} = {\partial_t}$ becomes null, which specifically means $\chi^i \chi_i = g_{tt} = 0$.

In practice, the larger of the two roots of the horizon structure is associated with the outer SLS, typically denoted as $r^+_{SLS}$. This is because the ergoregion lies between $r_+ < r < r^+_{SLS}$, at which point the time-like Killing vector $\chi^i$ transitions to space-like (see Fig. \ref{m600}). Thus, an observer must follow the world line of $\chi^i$ as shown in Fig. \ref{m600} for the RASN-BH metric function (\ref{RASN metric}). Upon closer inspection of the ergoregion behavior for the RASN-BH, it is observed that the ergoregions expand with an increase in the flavor number $N$ as well as the parameter $K$, implying that they are larger than those of the Kerr BH ($\Lambda = K = 0$). In contrast, an increase in the parameter $\kappa$ results in a shrinkage of the ergoregion's size (see Fig. \ref{m600}). 
\begin{figure}[h!]
 \begin{center}
 \subfigure[]{
 \includegraphics[height=4.1cm,width=4.1cm]{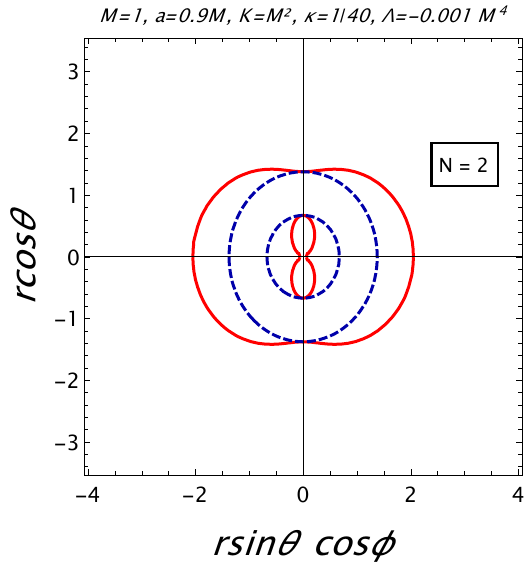}
 \label{600a}}
 \subfigure[]{
 \includegraphics[height=4.1cm,width=4.1cm]{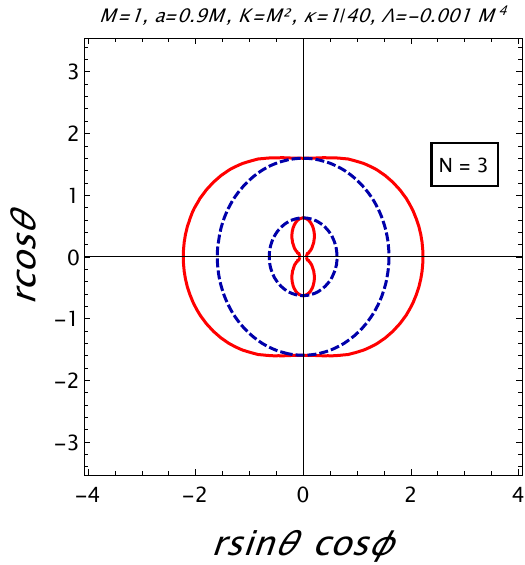}
 \label{600b}}
 \subfigure[]{
 \includegraphics[height=4.1cm,width=4.1cm]{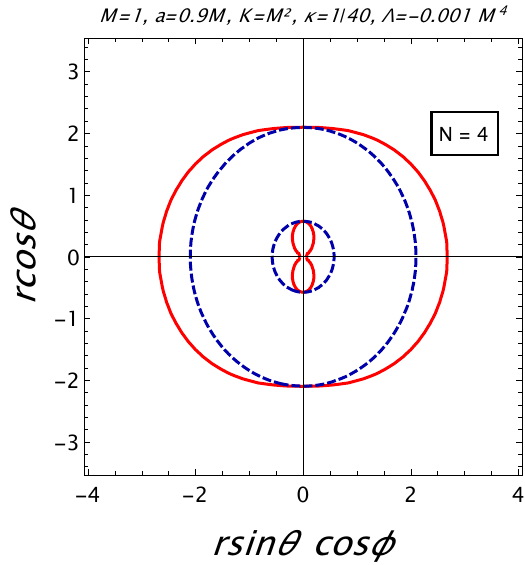}
 \label{600c}}
 \subfigure[]{
 \includegraphics[height=4.1cm,width=4.1cm]{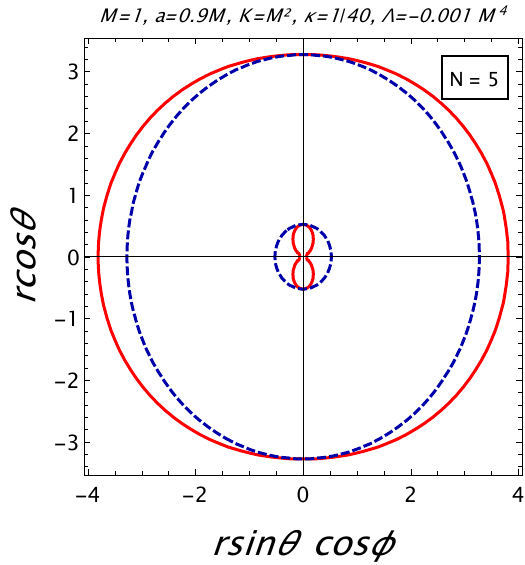}
 \label{600d}}\\
 \subfigure[]{
 \includegraphics[height=4.1cm,width=4.1cm]{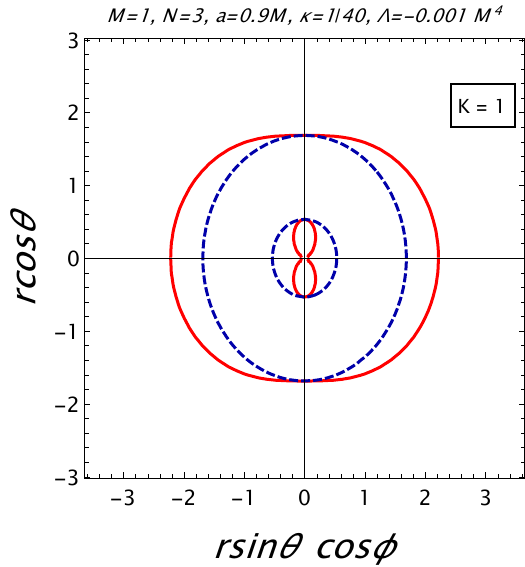}
 \label{600e}}
 \subfigure[]{
 \includegraphics[height=4.1cm,width=4.1cm]{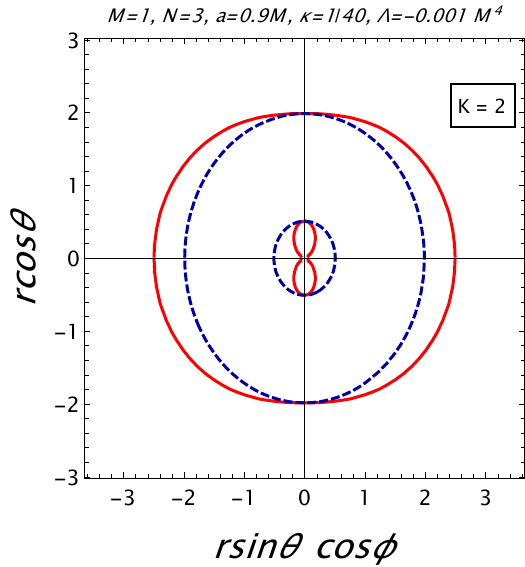}
 \label{600f}}
 \subfigure[]{
 \includegraphics[height=4.1cm,width=4.1cm]{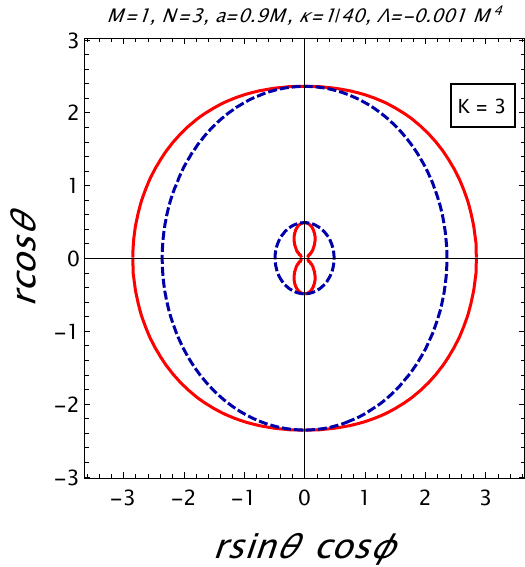}
 \label{600g}}
 \subfigure[]{
 \includegraphics[height=4.1cm,width=4.1cm]{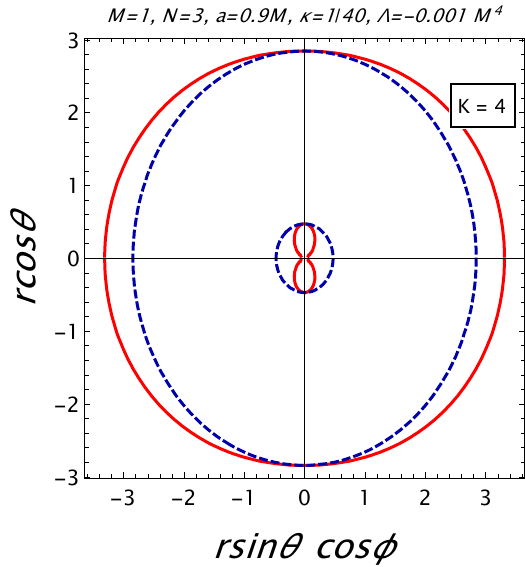}
 \label{600h}}\\
  \subfigure[]{
 \includegraphics[height=4.1cm,width=4.1cm]{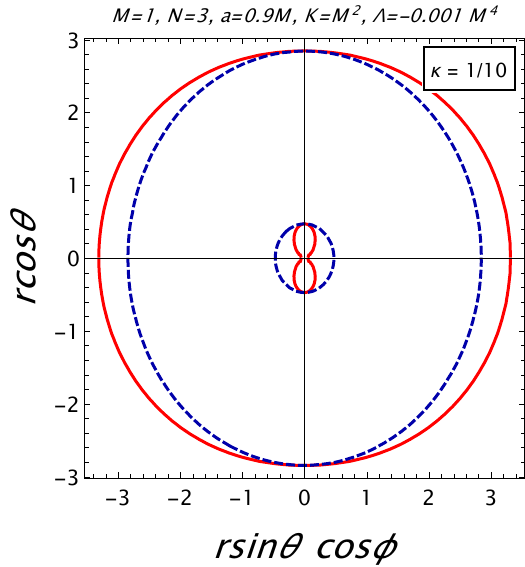}
 \label{600e}}
 \subfigure[]{
 \includegraphics[height=4.1cm,width=4.1cm]{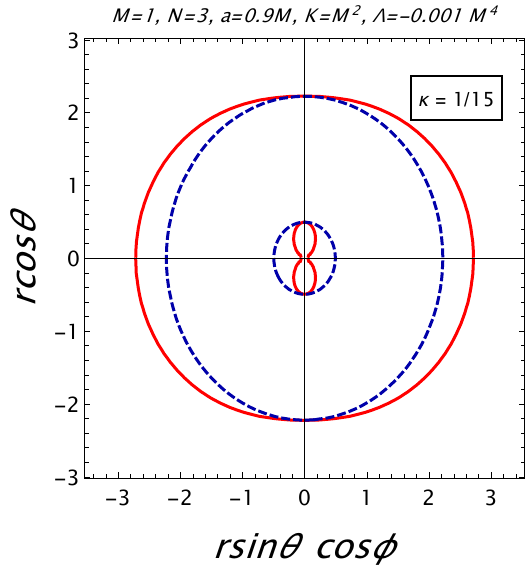}
 \label{600f}}
 \subfigure[]{
 \includegraphics[height=4.1cm,width=4.1cm]{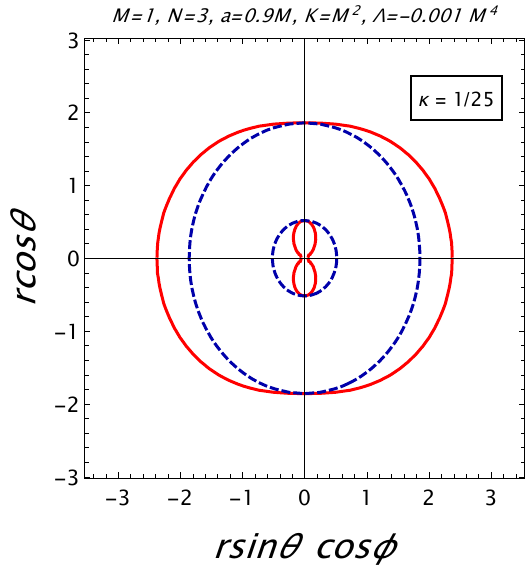}
 \label{600g}}
 \subfigure[]{
 \includegraphics[height=4.1cm,width=4.1cm]{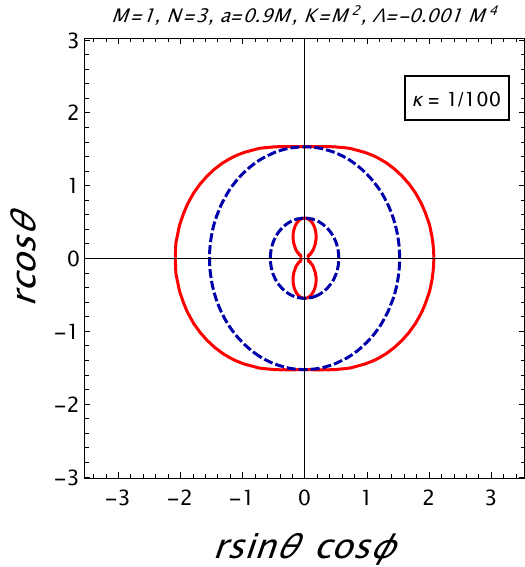}
 \label{600h}}\\
  \subfigure[]{
 \includegraphics[height=4.1cm,width=4.1cm]{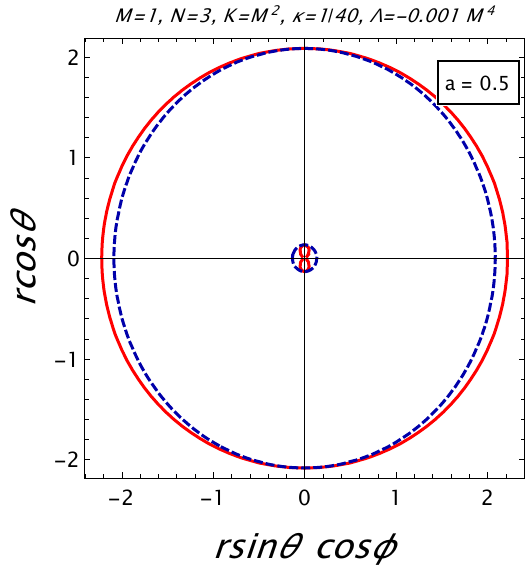}
 \label{600e}}
 \subfigure[]{
 \includegraphics[height=4.1cm,width=4.1cm]{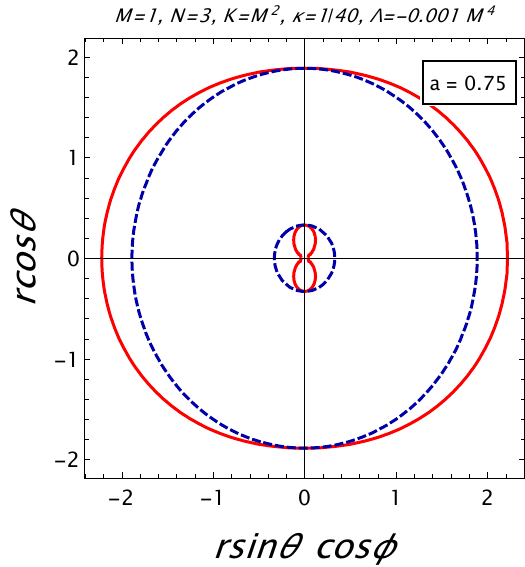}
 \label{600f}}
 \subfigure[]{
 \includegraphics[height=4.1cm,width=4.1cm]{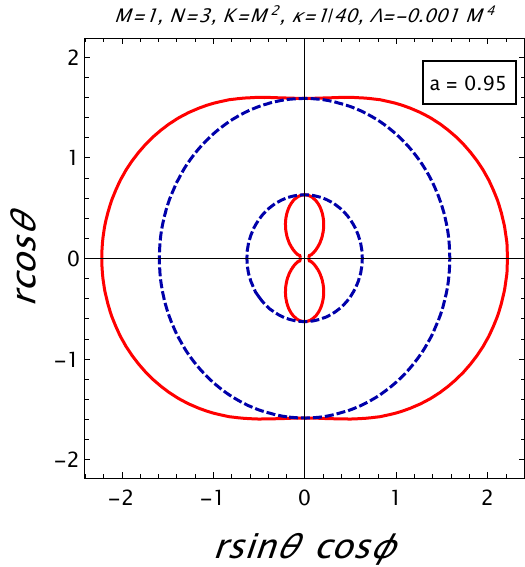}
 \label{600g}}
 \subfigure[]{
 \includegraphics[height=4.1cm,width=4.1cm]{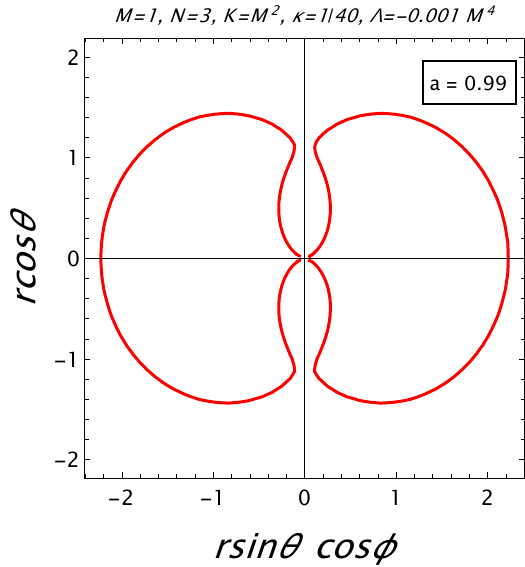}
 \label{600h}}
  \caption{The cross-section of the event horizon (outer blue curve), the SLS (outer red dotted curve), and the ergoregion of the RASN-BH for different values of the parameters $(N, K, \kappa, a)$.}
 \label{m600}
 \end{center}
 \end{figure}
Notably, the ergoregion is of key importance as energy can be extracted from this region via the Penrose process.

\section{Photon orbits and the BH shadow}\label{sec:photons}

In this section, we aim to provide a comprehensive study of shadow behavior involving the null geodesic formalism relevant to photon orbits around the RASN-BH. In this regard, we base our study on the common Lagrangian dynamics to obtain the equations of motion from the Hamilton-Jacobi equation. We then solve these equations numerically to obtain the BH shadow. In fact, the Hamilton-Jacobi equation is given by
\begin{equation} \label{e33}
    \frac{\partial S}{\partial \lambda}=-H,
\end{equation}
where $S$ is the Jacobi action, $\lambda$ is the affine parameter of the trajectory curves, and $x^{\alpha }$ indicates  the coordinates. The Hamiltonian is then given by
\begin{equation} \label{e34}
    H=\frac{1}{2}g^{\alpha \beta }\frac{
    \partial S}{\partial x^{\alpha}}\frac{\partial S}{\partial x^{\beta }},
\end{equation}
for geodesic curves in vacuum,  from which, Eq. \eqref{e33} yields
\begin{equation} \label{e35}
    \frac{\partial S}{\partial \lambda }=-\frac{1}{2}g^{\alpha \beta }\frac{
    \partial S}{\partial x^{\alpha }}\frac{\partial S}{\partial x^{\beta }}.
\end{equation}
Based on the four-momentum $p_\alpha = \partial S/\partial x_\alpha$, and using Carter's separability prescription \cite{Chandrasekhar:2002}, the action can be separated as
\begin{equation} \label{e36}
    S=\frac{1}{2}\mu ^{2}\lambda -Et+L\phi +S_{r}(r)+S_{\theta }(\theta),
\end{equation}
where $\mu$ stands for the rest mass of the particles. However, to consider photon orbits, we simply take $\mu=0$. The symmetry of the spacetime enables us to specify the associated conserved energy and angular momentum as follows:
\begin{eqnarray}
    E&=&-p_t=g_{tt}\dot{t}+g_{t\phi}\dot{\phi},\\
    L&=&p_\phi=g_{t\phi}\dot{t}+g_{\phi\phi}\dot{\phi}.
\end{eqnarray}
Upon implementing into account the RASN-BH's geometry, the equations of motion according to the four differential equations in the $r$-$\theta$ plane can be stated as \cite{Slany:2020jhs,PhysRevD.81.044020,StuchlikEPJC2018}
\begin{eqnarray} \label{eos}
    \rho^2\frac{\mathrm{d}t}{\mathrm{d}\lambda}&=&{\frac{a \Sigma^{2}}{\Delta_\theta}\left(L-a E\sin^2\theta\right)+\frac{(a^2+r^2)\Sigma^{2}}{\Delta_r}\Bigl[(a^2+r^2)E- a  L\Bigr]},\label{eq:tdot}\\
    \rho^2\frac{\mathrm{d}r}{\mathrm{d}\lambda}&=&\epsilon_r\sqrt{R(r)}, \label{eq:rdot} \\
    \rho^2\frac{\mathrm{d}\theta}{\mathrm{d}\lambda}&=&\epsilon_\theta\sqrt{\Theta(\theta)}, \label{eq:thetadot} \\
     \rho^2\frac{\mathrm{d}\phi}{\mathrm{d}\lambda}&=&{\frac{\Sigma^{2}}{\Delta_\theta}\left(L \csc^2\theta-a E\right)+\frac{a\Sigma^{2}}{\Delta_r}\Bigl[(a^2+r^2)E-a L\Bigr]},\label{eq:phodot}
\end{eqnarray}
in which $\epsilon_r =\epsilon_\theta=\pm1$, with
\begin{subequations}
\begin{align}
    & R(r) = {\Sigma^{2}\Bigl[\left(a^2+r^2\right)E- a  L\Bigr]^2-\Bigl[(L-aE)^2+\mathcal{O}\Bigr]\Delta_r,} \label{eq:R} \\
    & \Theta(\theta) = \Bigl[(L-aE)^2+\mathcal{O}\Bigr]\Delta_\theta+{\Sigma^{2}}{\cos^{2}\theta}\Bigl[a^2 E^2-L^2 \csc^2\theta\Bigr]. \label{eq:Theta}
    \end{align}
\end{subequations}
In the above expressions, $\mathcal{O} = \mathscr{Q} - (L-aE)^2$ refers to the generalized Carter's constant \cite{Carter:1968rr}, where $\mathscr{Q}$ is the Carter's separation constant. The photon trajectories around the RASN-BH are governed by the four equations \eqref{eq:tdot}-\eqref{eq:phodot}. These trajectories depend fundamentally on the impact parameters $\xi = L / E$ and $\eta = \mathscr{Q} / E^2$ \citep{Chandrasekhar:2002}. Broadly speaking, photons exhibit three different types of trajectories: scattering, spherical, and plunging orbits. Additionally, with respect to $\xi$ and $\eta$, the radial equations \eqref{eq:R} and \eqref{eq:Theta} take the following forms:
\begin{subequations}
\begin{align}
  &  R(r)={E^2}\Sigma^{2}\Bigl[ \Bigl((a^2+r^2)- a \xi\Bigr)^2-\Delta_r\Bigl((\xi-a)^2+\eta\Bigr)\Bigr],\label{eq:R(r)}\\
  &  \Theta(\theta) = E^2\Bigl[
  \eta\Delta_\theta+{\Sigma^{2}}{\cos^{2}\theta}\left(a^2-\xi^2\csc^2\theta\right)
  \Bigr].
    \end{align}
\end{subequations}
In the following sections, the study will focus on exploring the geometric structure of the photon regions and the appearance of the shadow in relation to the RASN-BH, utilizing relevant observable parameters.

\subsection{Orbits of constant radius and the photon regions}\label{subsec:photonregions}

In fact, the study of BH shadow relies strictly on photon orbits corresponding to paths where the radial coordinate remains constant, closely skimming the BH’s horizon. These orbits, known as spherical photon orbits, play a critical role in determining whether a photon escapes to infinity or plunges into the BH. Due to their inherent instability, spherical photon orbits form an infinite sequence of photon rings that delineate the BH's shadow. In static spacetimes, such as Schwarzschild, these orbits are confined to a single plane. However, in rotating BHs, the frame-dragging effect induces a photon region where spherical orbits become non-planar. This region, bounded by the innermost and outermost circular orbits, defines the domain of spherical photon orbits, as studied extensively in Kerr BH solutions (see Refs.~\cite{Chandrasekhar:2002,Bardeen:1972a,Bardeen:1973b}). Investigations into the properties and observational implications of these orbits have expanded significantly in Kerr and Kerr-like spacetimes (see Refs.~\cite{stoghianidis_polar_1987,cramer_using_1997,Teo:2003,Johannsen:2013,Grenzebach:2014,Perlick:2017,charbulak_spherical_2018,Johnson_universal_2020,Himwich:2020,Gelles:2021,Ayzenberg:2022,Das:2022,fathi_spherical_2023,ANJUM2023101195,Chen:2023,andaru_spherical_2023}).

Based on the formalism presented above, the spherical photon orbits at a given radius $r_p$ are determined by the criteria outlined in \cite{Teo:2003}:
\begin{equation} \label{e39}
    R(r_p) = R'(r_p) = 0,
\end{equation}
which corresponds to the radii of spherical photon orbits around the BH. By exploiting the form given in Eq. \eqref{eq:R(r)}, one obtains the following physically acceptable set of solutions:
\begin{eqnarray} 
    \xi_p &=& \left( \frac{\Delta_r'(r^{2}+a^{2}) - 4\Delta_r r}{a \Delta_r' \Sigma} \right)_{r_p}, \label{e46} \\
    \eta_p &=& \left( \frac{16r^{2} \Delta_r (a^{2} - \Delta_r) - r^{4} \Delta_r'^{2} + 8r^{3} \Delta_r \Delta_r'}{a^{2} \Delta_r'^{2} \Sigma^2} \right)_{r_p}, \label{e47}
\end{eqnarray}
within the photon region. Accordingly, spherical photon orbits confined to the equatorial plane are determined by the condition $\Theta(\pi/2) = 0$ (or equivalently, $\eta_p = 0$). The two largest positive roots of this equation correspond to the prograde radius, $r_p^-$, and retrograde radius, $r_p^+$, which define the innermost and outermost spherical photon orbits. In this context, $r_p^\pm$ mark the inner and outer boundaries of the photon region. 
The photon region is further characterized by the condition $\Theta(\theta) \geq 0$ for spherical photon orbits. By using the critical impact parameters in Eqs. \eqref{e46} and \eqref{e47} in the angular potential \eqref{eq:Theta}, we can solve for the radii with $\eta_p = 0$. The resulting photon regions around the RASN-BH are illustrated in Fig. \ref{fig:photonRegion} for various values of the BH's parameters.
\begin{figure}[h]
    \centering
    \includegraphics[width=5.4cm]{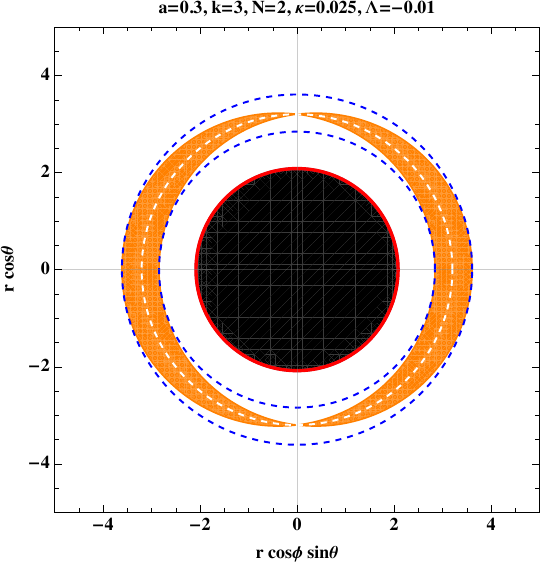} (a)
    \includegraphics[width=5.4cm]{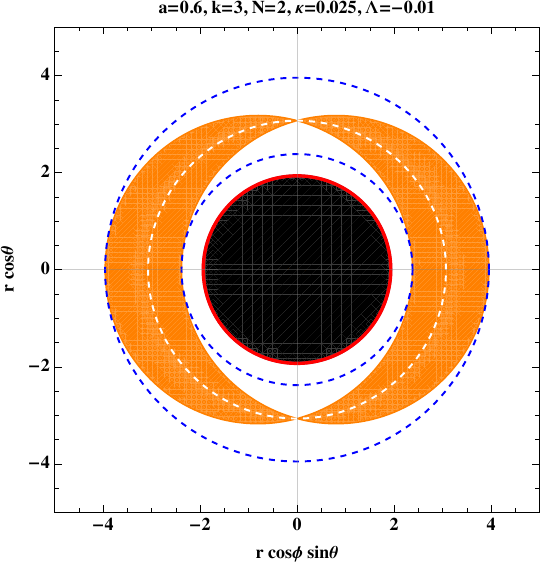} (b)
     \includegraphics[width=5.4cm]{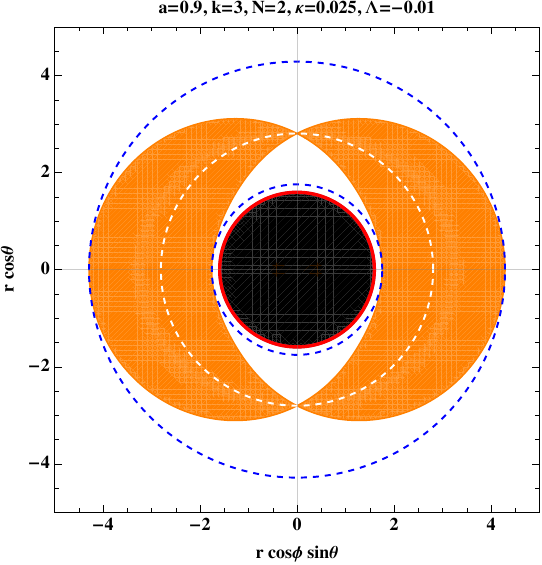} (c)
     \includegraphics[width=5.4cm]{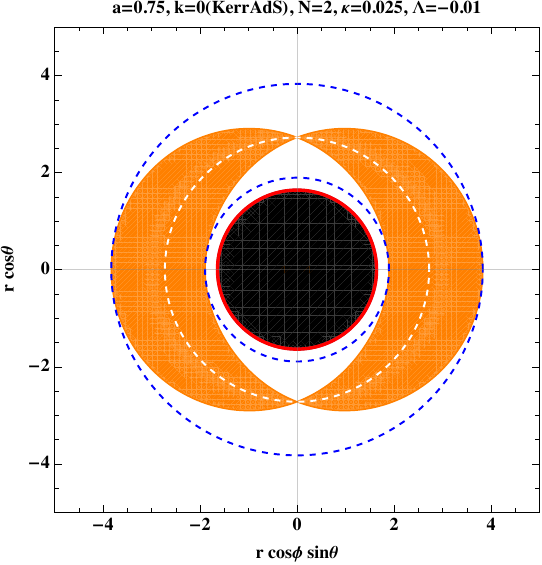} (d)
     \includegraphics[width=5.4cm]{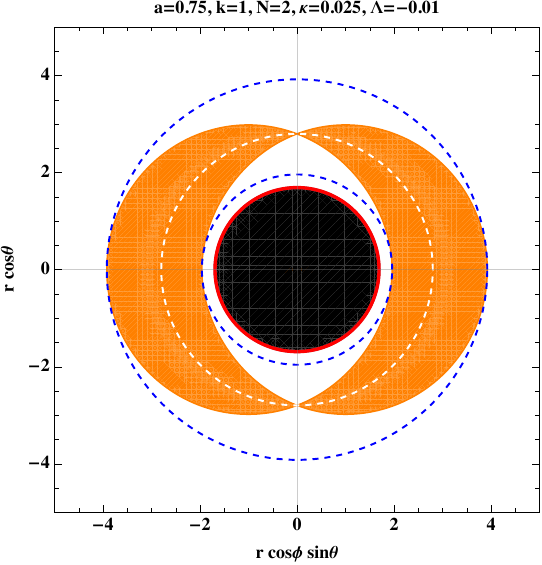} (e)
     \includegraphics[width=5.4cm]{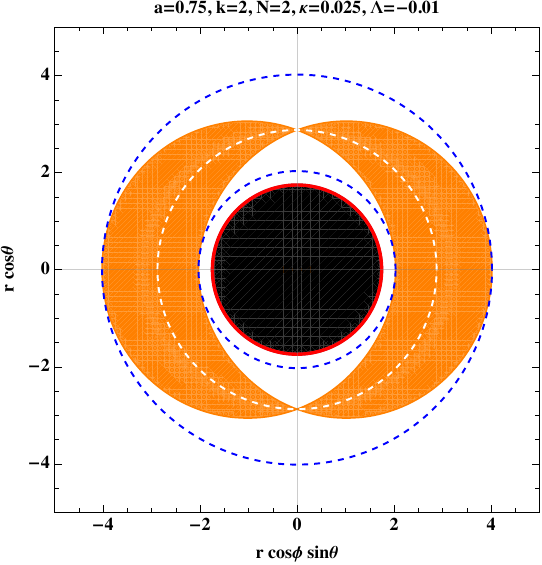} (f)
    \caption{The photon regions around the RASN-BH in the polar plane (indicated by cyan areas) plotted for various values of the spin parameter and the $K$-parameter, assuming $\kappa = 0.025$ and $\Lambda = -0.01$. The blue dashed circles represent the radii of exterior planar photon orbits, denoted as $r_{p}^\pm$. The white dashed circles correspond to the radius of $r_0$, while the BH (black region) has the radius of $r_+$. The panels display the following values: (a) $r_+ = 2.082$, $r_{p}^- = 2.843$, $r_{p}^+ = 3.609$, and $r_0 = 3.202$; (b) $r_+ = 1.932$, $r_{p}^- = 2.381$, $r_{p_+} = 3.955$, and $r_0 = 3.070$; (c) $r_+ = 1.593$, $r_{p}^- = 1.757$, $r_{p}^+ = 4.287$, and $r_0 = 2.723$; 
    (d) $r_+ = 1.639$, $r_{p}^- = 1.894$, $r_{p_+} = 3.828$, and $r_0 = 3.070$; 
    (e) $r_+ = 1.690$, $r_{p}^- = 1.961$, $r_{p_+} = 3.921$, and $r_0 = 2.80$;
    (f) $r_+ = 1.744$, $r_{p}^- = 2.030$, $r_{p_+} = 4.019$, and $r_0 = 2.880$;
     The units along the axes are given in terms of $M$.}
    \label{fig:photonRegion}
\end{figure}
In these diagrams, in addition to the radii of planar orbits, we also show the radii of polar orbits, $r_0$, which represent the spherical orbits for photons with no associated angular momentum. These photons traverse the entire polar angle, passing through the poles without any change in the azimuthal angle. The radii of such orbits are obtained by solving the equation $\xi_p = 0$, using the expression in Eq. \eqref{e46}. From the diagrams, it can be inferred that as the spin parameter increases while $K$ remains constant, the BH shrinks in size, whereas the photon region expands. Conversely, for a fixed spin parameter, a larger $K$ results in a bigger BH, and although the shape of the photon region remains unchanged, it becomes enlarged as $K$ increases. Notably, the RASN-BH is larger than the Kerr-AdS BH (Fig. \ref{fig:photonRegion}(d)), both in terms of the BH size and the photon regions.

It is important to note that for a distant observer, the photons residing within the photon regions (i.e., photons on spherical orbits) form the innermost photon rings, as they complete numerous half-orbits around the BH before reaching the observer (for further discussion, see Refs. \cite{Gralla:2019}, \cite{bisnovatyi-kogan_analytical_2022}, \cite{tsupko_shape_2022}, as well as the foundational works in Refs. \cite{claudel_geometry_2001}, \cite{virbhadra_relativistic_2009}, \cite{virbhadra_compactness_2024}). These photons thus play a key role in defining the true boundary of the shadow (or the critical curve), which will be explored in the next subsection.

\subsection{The BH shadow}

Escaping photons in the unstable orbit give the possibility for distant observers to backward-trace and obtain a shadow cast for an observer located at the spatial position $(r_o, \theta_o)$. Such an observer is also known as the Zero Angular Momentum Observers (ZAMOs), for whom the shadow is seen in the celestial plane, identified by the coordinates \cite{Johannsen:2013vgc}
\begin{eqnarray} \label{e48}
    X  &=& -r_{o} \frac{\xi }{\zeta \sqrt{g_{\phi \phi }} \left( 1 + \frac{g_{t\phi }}{g_{\phi \phi }} \xi \right)}, \\
    Y  &=& r_{o} \frac{\pm \sqrt{\Theta (\theta_o)}}{\zeta \sqrt{g_{\theta \theta }} \left( 1 + \frac{g_{t\phi}}{g_{\phi \phi }} \xi \right)},
\end{eqnarray}
where
\begin{equation}
\zeta = \sqrt{\frac{g_{\phi\phi}}{g_{t\phi}^2 - g_{tt} g_{\phi\phi}}}.
    \label{eq:zeta}
\end{equation}
In the limit $r_{o} \rightarrow \infty$, these coordinates simplify to
\begin{eqnarray} \label{e49}
    X &=& -\xi \csc \theta_o,   \label{eq:X} \\
    Y &=& \pm \sqrt{\eta + a^2 \cos^2 \theta_o - \xi^2 \cot^2 \theta_o}\,. \label{eq:Y}
\end{eqnarray}
These expressions further simplify to $X = -\xi$ and $Y = \pm \sqrt{\eta}$ when $\theta_o = \pi/2$. If, additionally, $a = 0$ and $K = 0$, the shadow cast by the Schwarzschild-AdS BH (a circle) is obtained. To conduct an appropriate analysis based on the shadow's behavior, Fig. \ref{Sh1} shows two-dimensional contour plots illustrating the boundary of the shadow of the RASN-BH in the $X$-$Y$ plane.
\begin{figure}[h!]
	\centering
	\includegraphics[width=7.1cm,height=7cm]{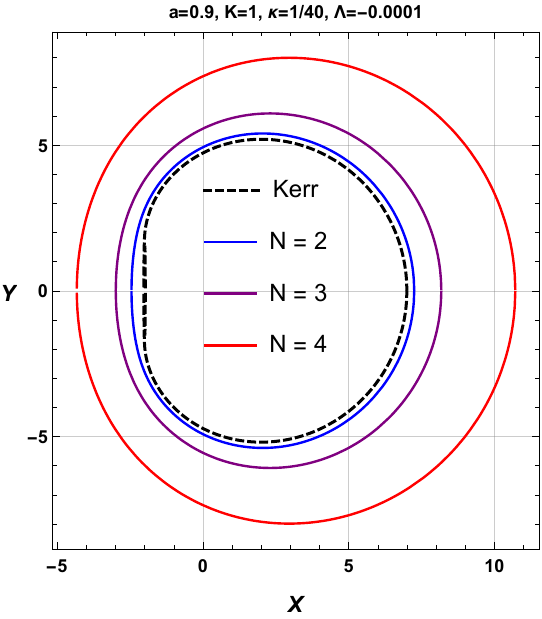}~~~(a)
	\includegraphics[width=7.1cm,height=7cm]{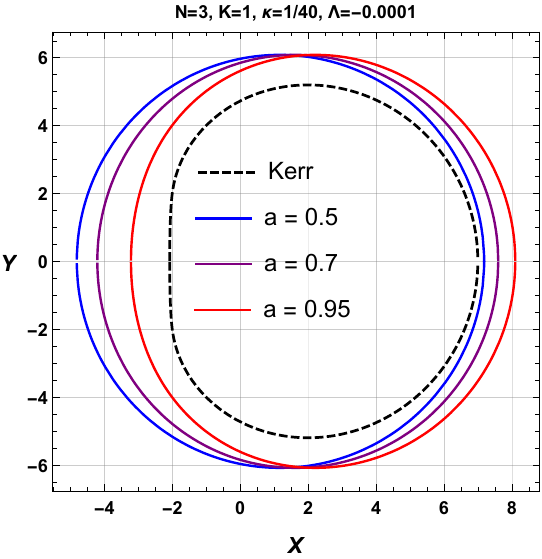}~~~~(b)\\
 \includegraphics[width=7.1cm,height=7cm]{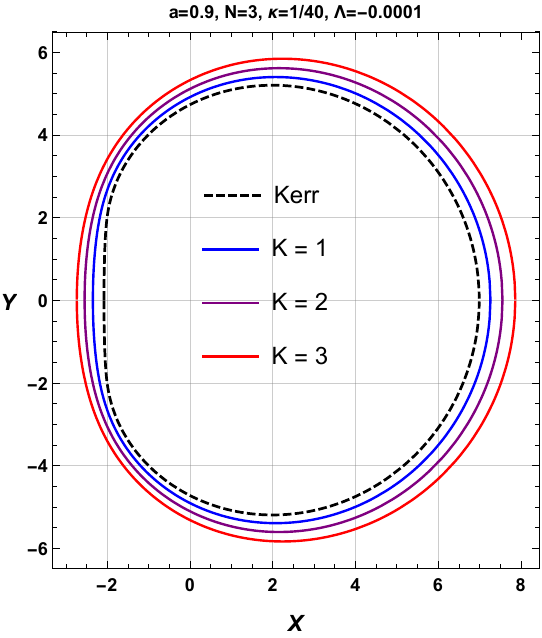}~~~(c)
\includegraphics[width=7.1cm,height=7cm]{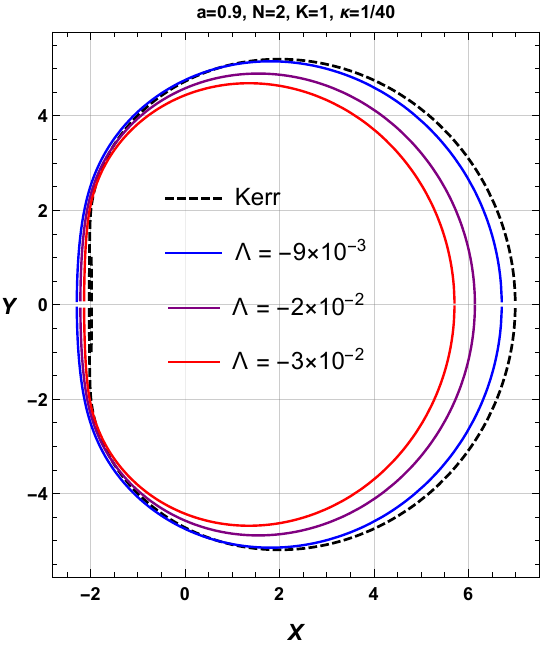}~~~~(d)
	\caption{Shadow silhouettes of RASN-BHs with varying all the parameter space for an inclination angle $\theta_o = \pi/2$. The unit of length along the axes is chosen to be the BH mass $M$.}
	\label{Sh1}
\end{figure}
In the essence of the Einstein-AdS-$\mathrm{SU}(N)$-NLSM theory, the shadow visualization therefore must be influenced by the flavor number $N$ based on observations from Fig. \ref{Sh1}(a). Notably, from a geometrical perspective, the size of the BH shadow increases as the flavor number $N$ grows. This could serve as evidence that the flavor number $N$ offers an allowed constraint region within the M87$^\star$ and Sgr A$^\star$ data, which will be discussed later. On the other hand, the spin parameter $a$ affects the size of the BH's shadow, implying what is known as the D-shaped topological nature. Varying the spin parameter $a$ in an increasing sense causes the D-shaped visualization to become visible, which implies, in other words, that the size of the black shadow is slowly shrinking (see Fig. \ref{Sh1}(b)). Inspecting the impact of the parameter $K$ on the shadow behaviors is clearly shown in Fig. \ref{Sh1}(c). So, varying $K$ in an increasing sense leads to raising the BH shadow size across deformed two-dimensional circles designed for the photon spheres. In addition, the variation in the cosmological constant offers a different topological shape to that previously discovered. In this case, and as observed in Fig. \ref{Sh1}(d), the decreases in $\Lambda$ have caused the shadow to shift to the right. This contrasts with the behavior of the shadow of equatorial circular photon orbits, namely that the right-hand side of the shadow shifts to the right as $\Lambda$ decreases. To illustrate the impact of the inclination angle on the behavior of the shadow, particularly from a geometric point of view, Fig. \ref{Sh2} shows the corresponding plots. 
\begin{figure}[h!]
	\centering
	\includegraphics[width=8cm,height=6cm]{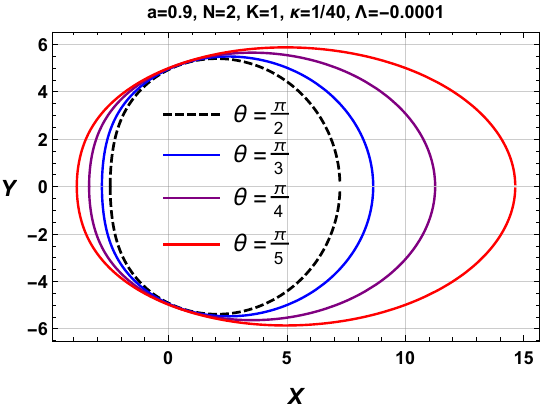}(a)
	\includegraphics[width=8cm,height=6cm]{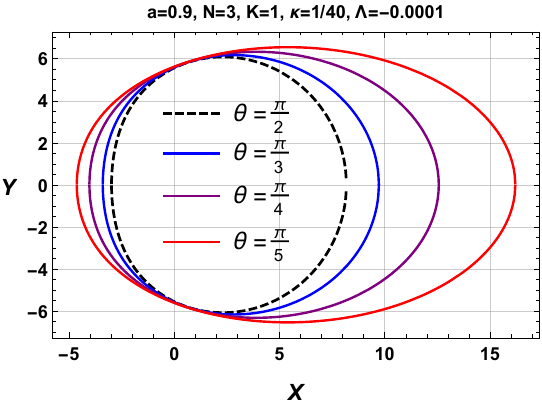}(b)\\
 \includegraphics[width=8cm,height=6cm]{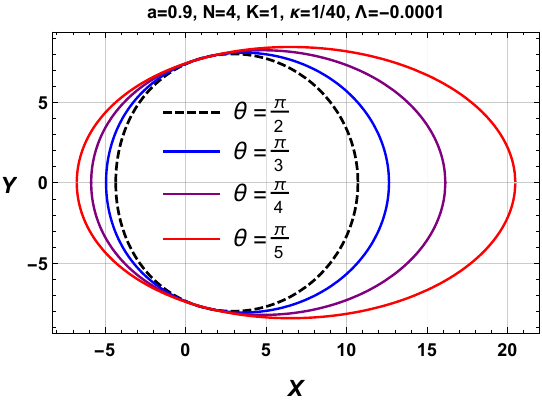}(c)
\includegraphics[width=8cm,height=6cm]{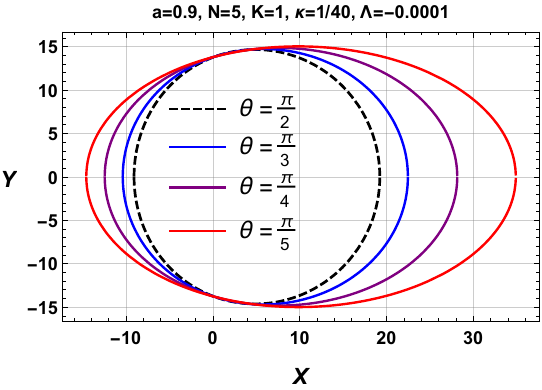}(d)
	\caption{Shadow silhouettes of the RASN BHs for different sets of inclination angles. The unit of length along the axes is chosen to be the BH mass $M$.}
	\label{Sh2}
\end{figure}
It is evident that a decrease in the inclination angle $\theta_o$ leads to an increase in the size of the BH shadow. Additionally, as expected, an increase in the flavor number $N$ also results in the enlargement of the shadow, as demonstrated in the diagrams (a)--(d) of Fig. \ref{Sh2}. {It is worth mentioning that a larger shadow may indicate a weaker gravitational field near the black hole's horizon, potentially suggesting gravitational stability. However, without more detailed studies on the black hole's response to perturbations, it is not possible to make definitive statements regarding its stability.}

\section{Shadow observables and BH parameter estimation}\label{sec:observables}

To have BH models that are observationally consistent, the parameters associated with theoretical BHs need to be constrained using EHT observations. The observed images of the supermassive BH M87* and Sgr A* are consistent with the predictions of the Kerr BH of general relativity. In the following, we identify key observables and examine their behavior within the framework of the RASN-BH. Note that several studies have explored constraints on the parameters of general black holes derived from the NJA (see, for example, Refs.~\cite{tsukamoto_constraining_2014,tsukamoto_black_2018}). In this section, we focus on the most salient aspects of the black hole, whose constraints provide valuable insights into its fundamental characteristics.

\subsection{Distortion}

In fact, the distortion measures the deviation of the BH shadow's shape from a circular reference. In other words, it quantifies the extent to which the shadow of a rotating BH departs from circularity. Here, we adopt the method proposed in Ref. \cite{Hioki:2009na}, where the linear radius of the shadow is defined as
\begin{eqnarray}
    R_s=\frac{(X_t-X_r)^2+Y_{t}^2}{2|X_r-X_t|},
\end{eqnarray}\label{Rs}
which is the radius of a hypothetical circle that tangentially touches the shadow at the points $(X_t, Y_t)$, $(X_b, Y_b)$, and $(X_r, 0)$ is defined in terms of the shadow coordinates $X$ and $Y$ as given in Eqs.~\eqref{eq:X} and \eqref{eq:Y}. Here, the subscripts $t$, $b$, and $r$ denote the top, bottom, and rightmost points, respectively, where the shadow intersects with the hypothetical circle. In this context, the distortion parameter can be expressed as
\begin{eqnarray}
    \delta_s = \frac{|X_l - X'_l|}{R_s},
\end{eqnarray}\label{deltas}
where $X_l$ and $X'_l$ correspond to the leftmost boundaries of the shadow and the reference circle, respectively, and lie along the $-X$-axis. For simplicity, assuming the shadow exhibits symmetry with respect to the $X$-axis, we set $X_t = X_b = 0$ and $Y_b = -Y_t$. Consequently, the point $Y_t \equiv Y(r_t)$ is determined by the condition $Y'(r_t) / X'(r_t) = 0$, with $r_t$ being the root of this equation. Similarly, $X_r \equiv X(r_r)$ and $X_l \equiv X(r_l)$ are identified as the two positive real roots of the equation $Y^2(r) = 0$. Figure~\ref{fig:deltas} shows the behavior of the distortion parameter $\delta_s$ for various illustrative cases.
\begin{figure}[h]
    \centering
    \includegraphics[width=7cm]{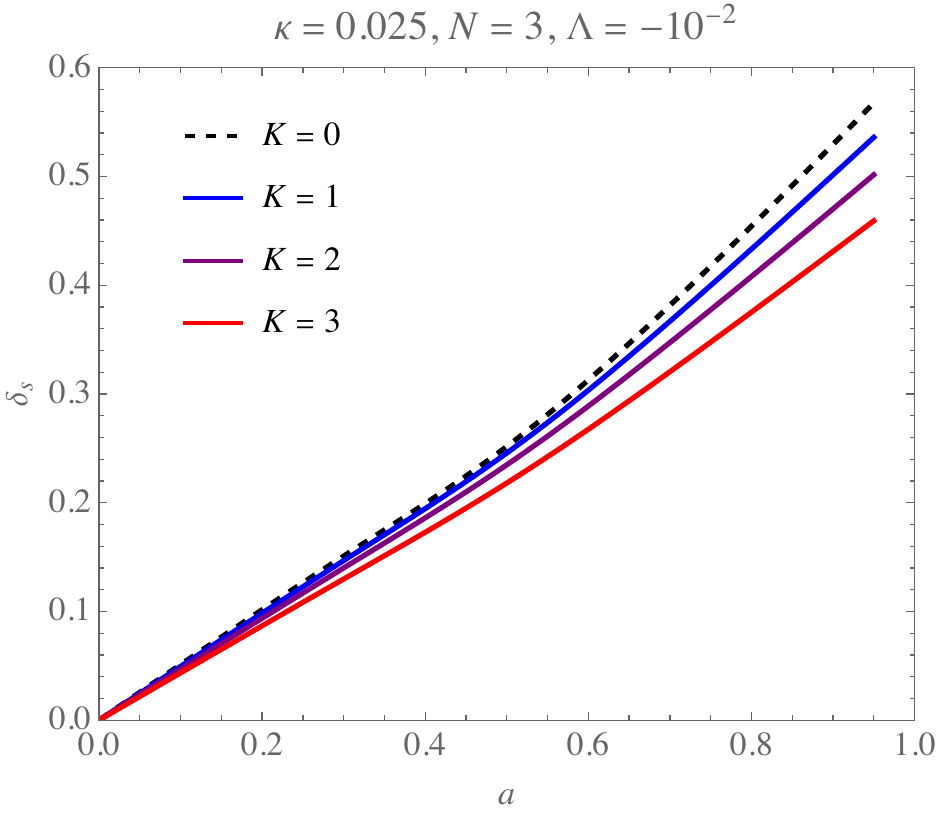} (a)
    \includegraphics[width=7cm]{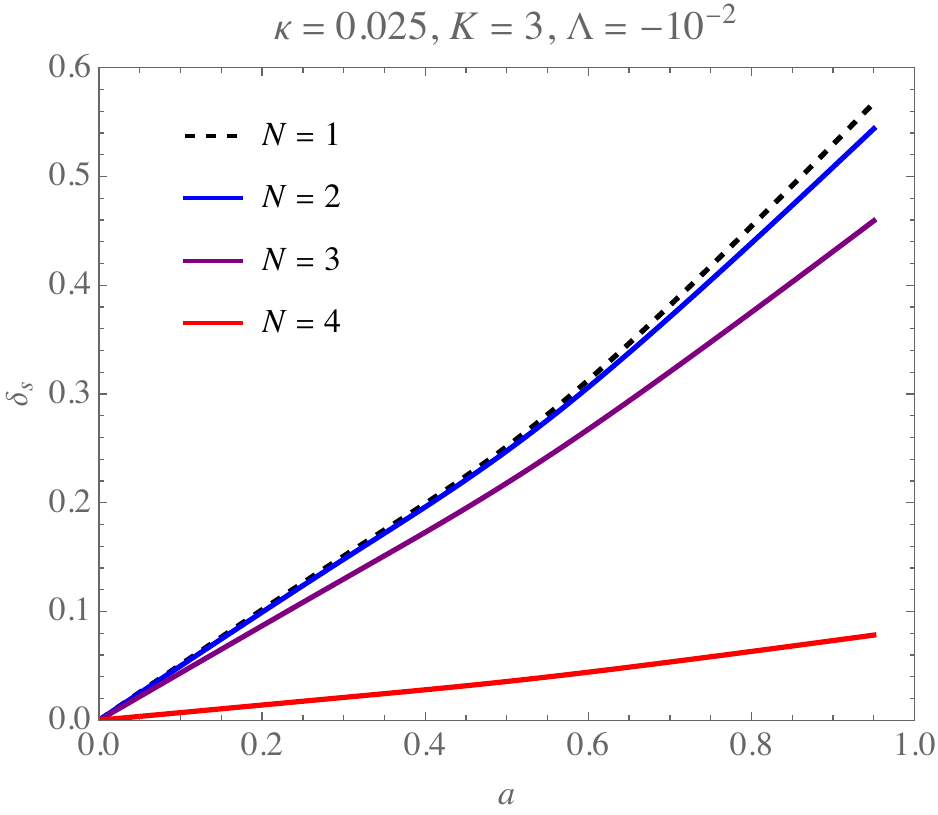} (b)
     \includegraphics[width=7cm]{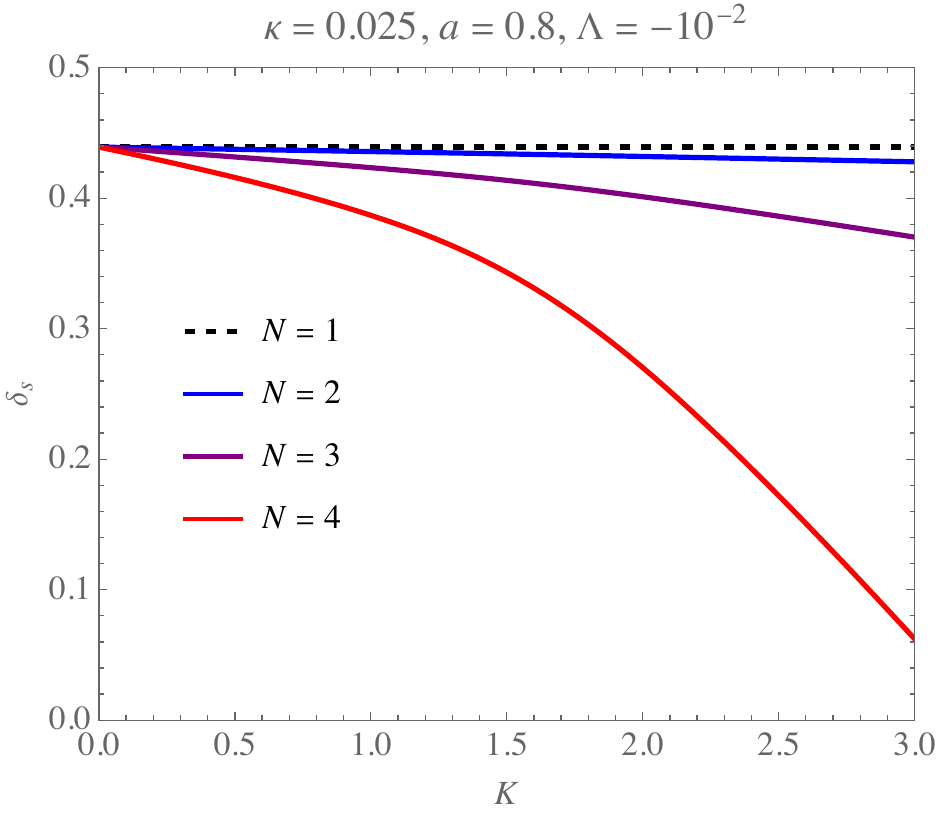} (c)
     \includegraphics[width=7cm]{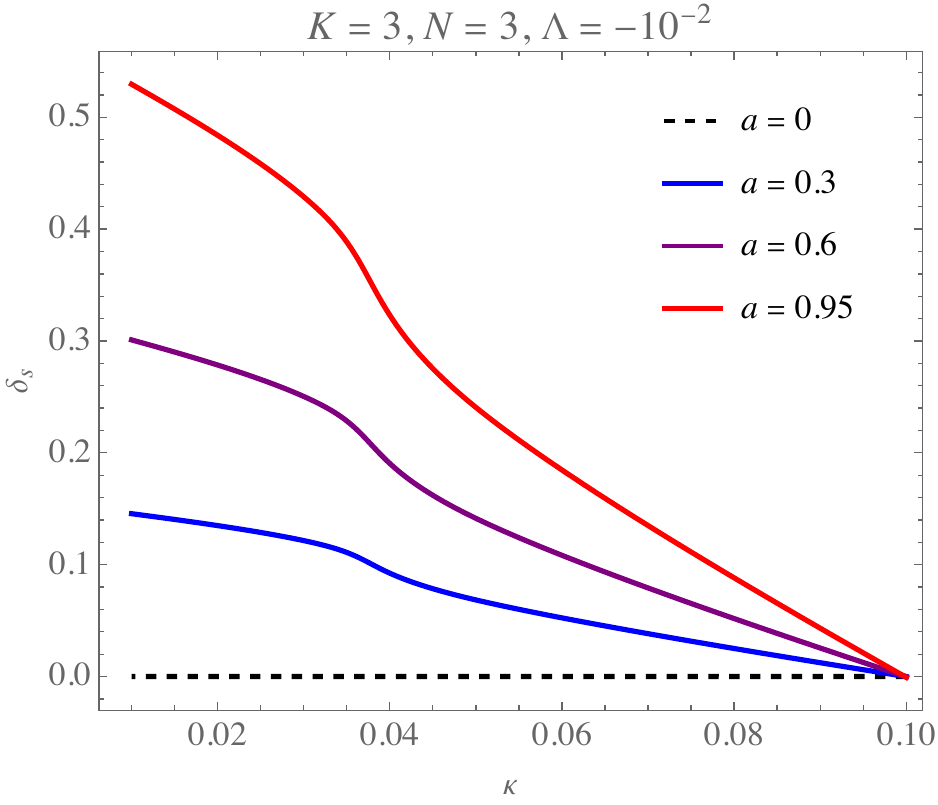} (d)
    \caption{The behavior of the distortion parameter $\delta_s$ is plotted for $\theta_o = \pi/4$, illustrating the variations with changes in different BH parameters. In panel (c), the convergent point for $K = 0$ is $\delta_s \approx 0.44$, while in panel (d), $\kappa_0 \approx 0.1$. The unit of length is taken to be the BH mass $M$.}
    \label{fig:deltas}
\end{figure}
Based on the diagrams, the variation of $\delta_s$ with respect to the spin parameter $a$, while $K$ and $N$ are fixed, exhibits an increasing trend. In contrast, an increase in either $K$ or $N$ results in a decrease in the distortion parameter, although the profiles remain increasing. Furthermore, for a fixed spin parameter, an increase in the $K$ parameter, with $N$ held constant, produces a descending profile for the distortion parameter. Similarly, increasing the $N$ parameter also reduces the distortion. However, all profiles converge to a single value at $K = 0$, which coincides with the value for $N = 1$. Finally, fixing $K$ and $N$, we observe that an increase in $\kappa$ causes the profiles to decrease until they reach $\delta_s = 0$ (corresponding to $a = 0$) at a specific value $\kappa_0$. Beyond this point, the profiles take on negative values. This transition indicates that for $\kappa > \kappa_0$, the BH shadows shift from being oblate to prolate for any fixed spin parameter $a \neq 0$.

\subsection{Parameter estimation}

The shadow of a BH reveals key characteristics of the underlying spacetime, such as its shape and size. As such, it provides a valuable tool for testing new gravity theories and for constraining BH parameters \citep{jusufi_constraints_2022,okyay_nonlinear_2022,afrin_parameter_2021,chen_superradiant_2022,afrin_tests_2023,sarikulov_shadow_2022,zahid_shadow_2023,zubair_4d_2023}. Estimations of BH parameters can be derived from the observables associated with the shadow. To achieve this, it is necessary to define the key parameters that describe the shadow's size and shape. 
In addition to the previously discussed method, a more direct approach can be employed, utilizing a coordinate-independent formalism \citep{abdujabbarov_coordinate-independent_2015,kumar_black_2020} and focusing on shadow observables such as the shadow's area and oblateness. The area of the BH shadow, denoted $A_s$, and its oblateness, $D_s$, provide insights into the deformation of the shadow and are defined as \citep{kumar_black_2020}
\begin{eqnarray}
A_s &=& 2\int_{r_p^{-}}^{r_p^+} Y(r_p) X'(r_p)\, \mathrm{d}r_p, \label{Area1} \\
D_s &=& \frac{Y_t - Y_b}{X_r - X_l} = \frac{2Y_t}{X_r - X_l}. \label{eq:Ds}
\end{eqnarray}
In Figs. \ref{fig:As} and \ref{fig:Ds}, we have plotted the sensitivity of the shadow area and deformation to the BH parameters.
\begin{figure}[h]
    \centering
    \includegraphics[width=7cm]{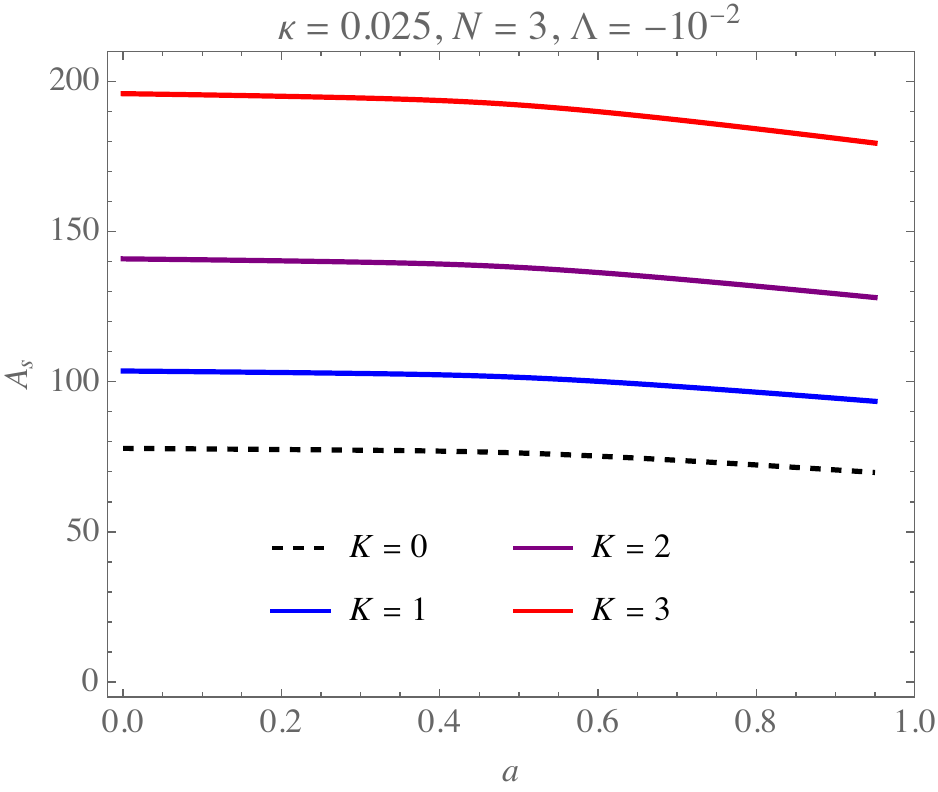} (a)
    \includegraphics[width=7cm]{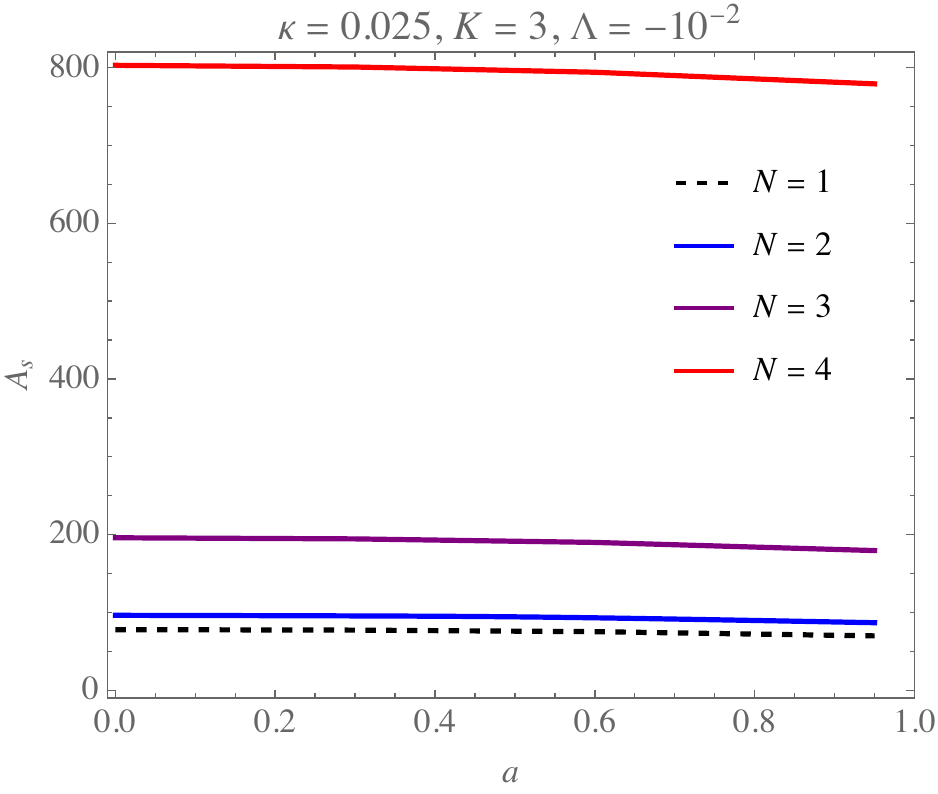} (b)
     \includegraphics[width=7cm]{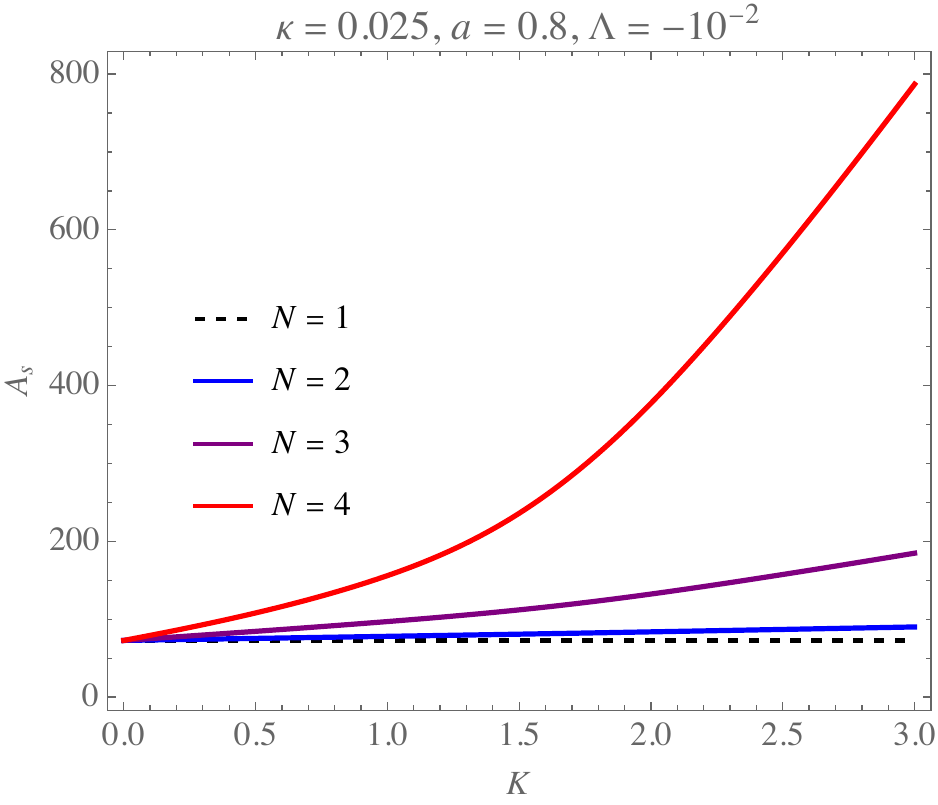} (c)
     \includegraphics[width=7cm]{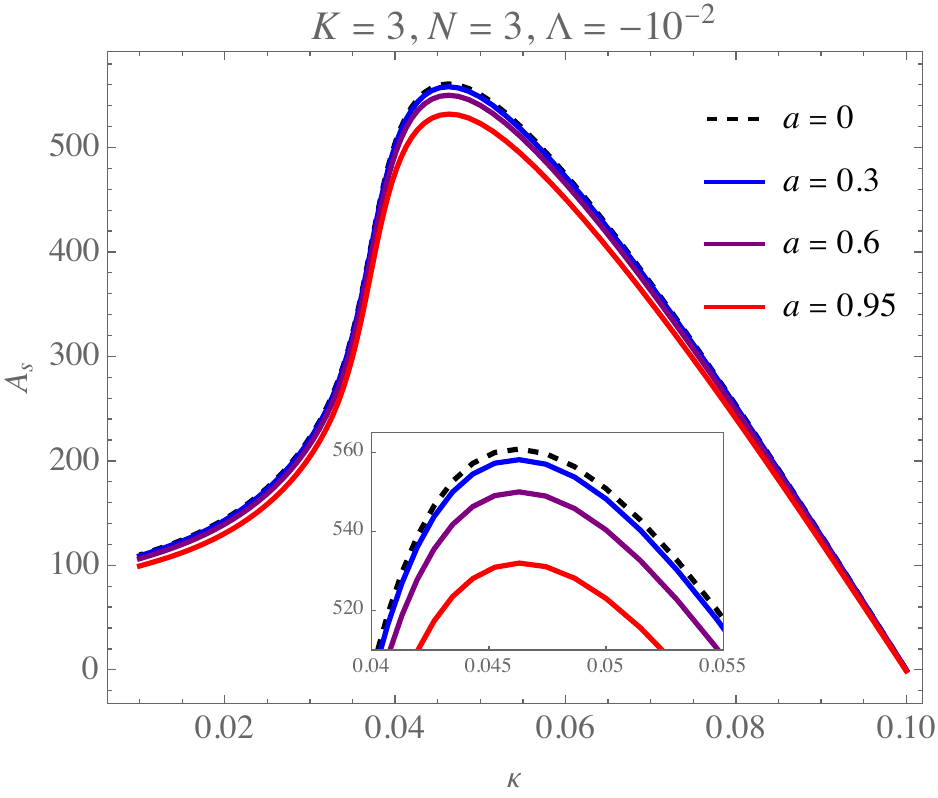} (d)
    \caption{The behavior of the shadow area $A_s$ as a function of various changes in BH parameters is shown, based on the initial conditions provided in Fig. \ref{fig:deltas}. The unit of length has been taken as the BH mass $M$.}
    \label{fig:As}
\end{figure}
\begin{figure}[h]
    \centering
    \includegraphics[width=7cm]{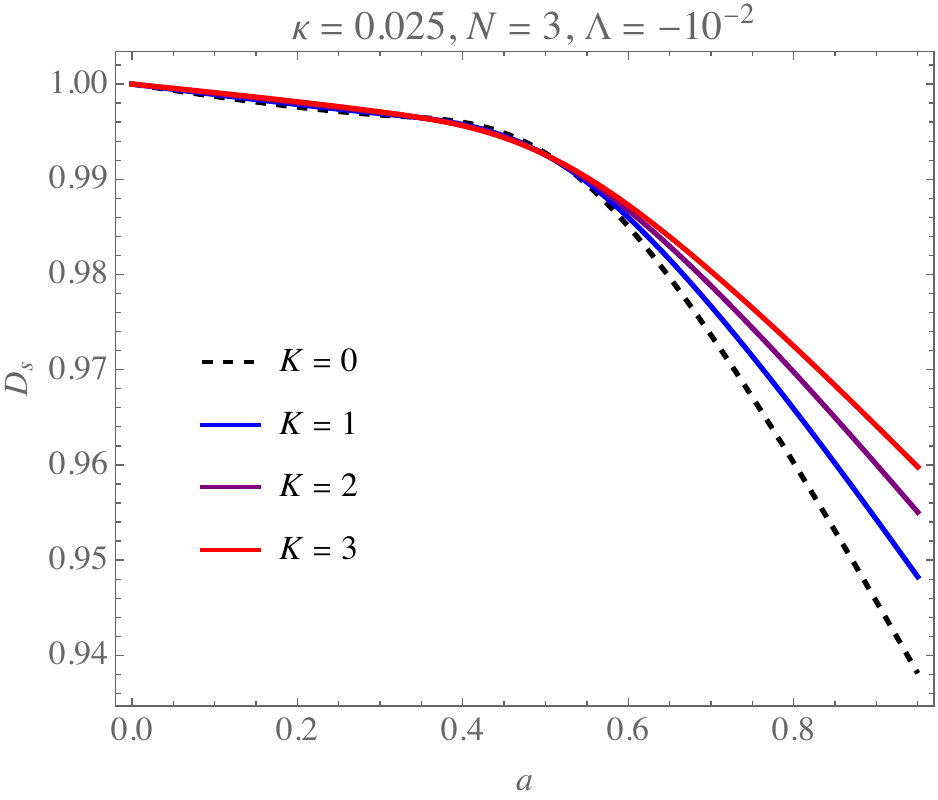} (a)
    \includegraphics[width=7cm]{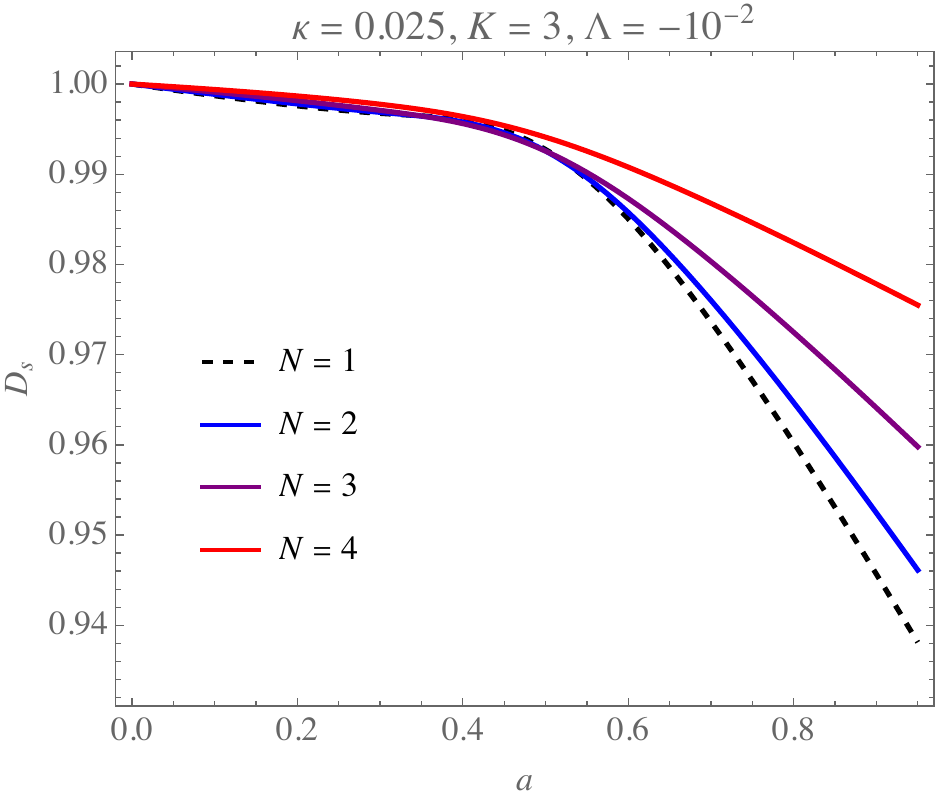} (b)
     \includegraphics[width=7cm]{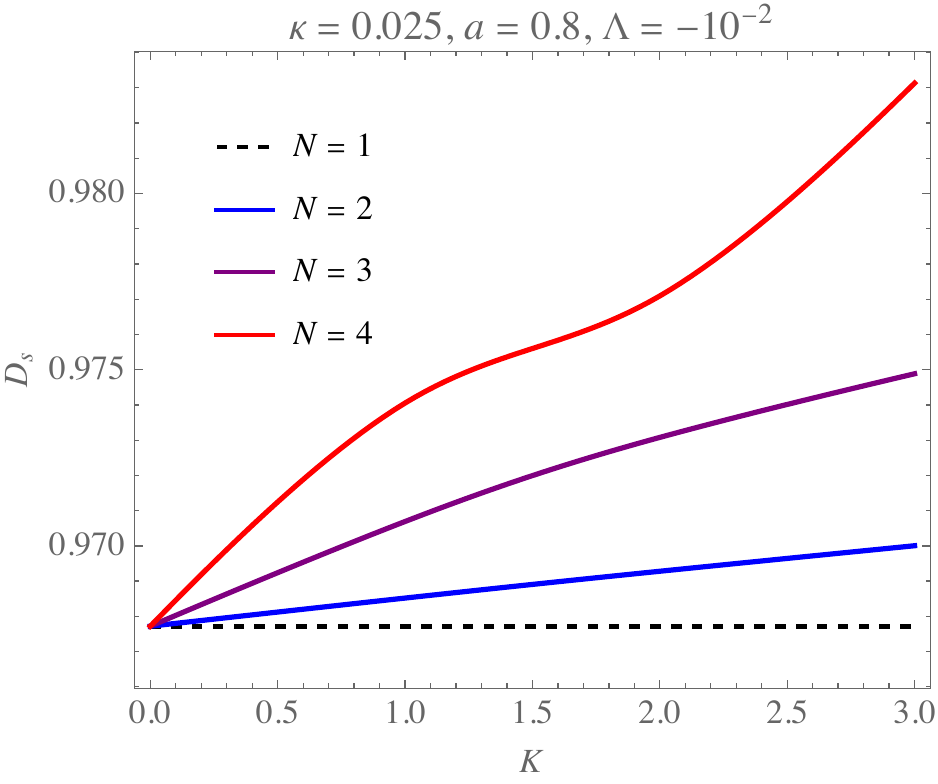} (c)
     \includegraphics[width=7cm]{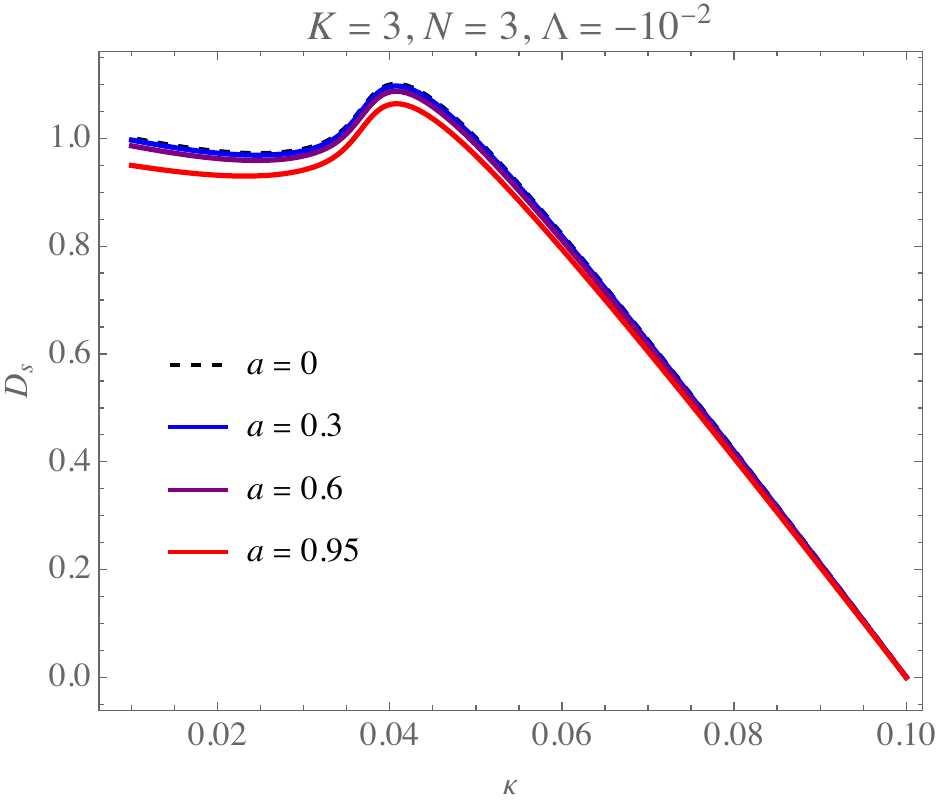} (d)
    \caption{The behavior of the deformation parameter $D_s$ as a function of various changes in BH parameters is shown, based on the initial conditions provided in Fig. \ref{fig:deltas}. The unit of length has been taken as the BH mass $M$.}
    \label{fig:Ds}
\end{figure}
It can be observed that an increase in the spin parameter reduces the shadow area and enhances the oblateness (i.e., a decrease in $D_s$). However, when $N$ is fixed, an increase in the $K$-parameter significantly enlarges the area of the oblate shadow, while simultaneously decreasing its oblateness. A similar behavior is observed with an increase in the $N$-parameter while keeping $K$ fixed, although the impact of $N$ on the oblate shadow area is more pronounced. Furthermore, for a fixed spin parameter, an increase in the $K$-parameter results in an enlarged shadow area, whereas an increase in $N$ under these conditions reduces the oblateness. Finally, when both $K$ and $N$ are fixed, increasing $\kappa$ leads to an initial growth in the shadow area, reaching a maximum before sharply declining until the shadow vanishes completely. During this process, the shadow's oblateness undergoes a slight increase, culminating in a fully prolate shadow at the maximum area. Subsequently, the shadow gradually becomes more oblate as it diminishes, eventually reducing to a straight line with zero width. Furthermore, in Fig. \ref{fig:AsDs}, we have demonstrated the mutual behaviors of $a$, $K$, and $N$ for fixed values of other parameters.
\begin{figure}[h]
    \centering
    \includegraphics[width=7cm]{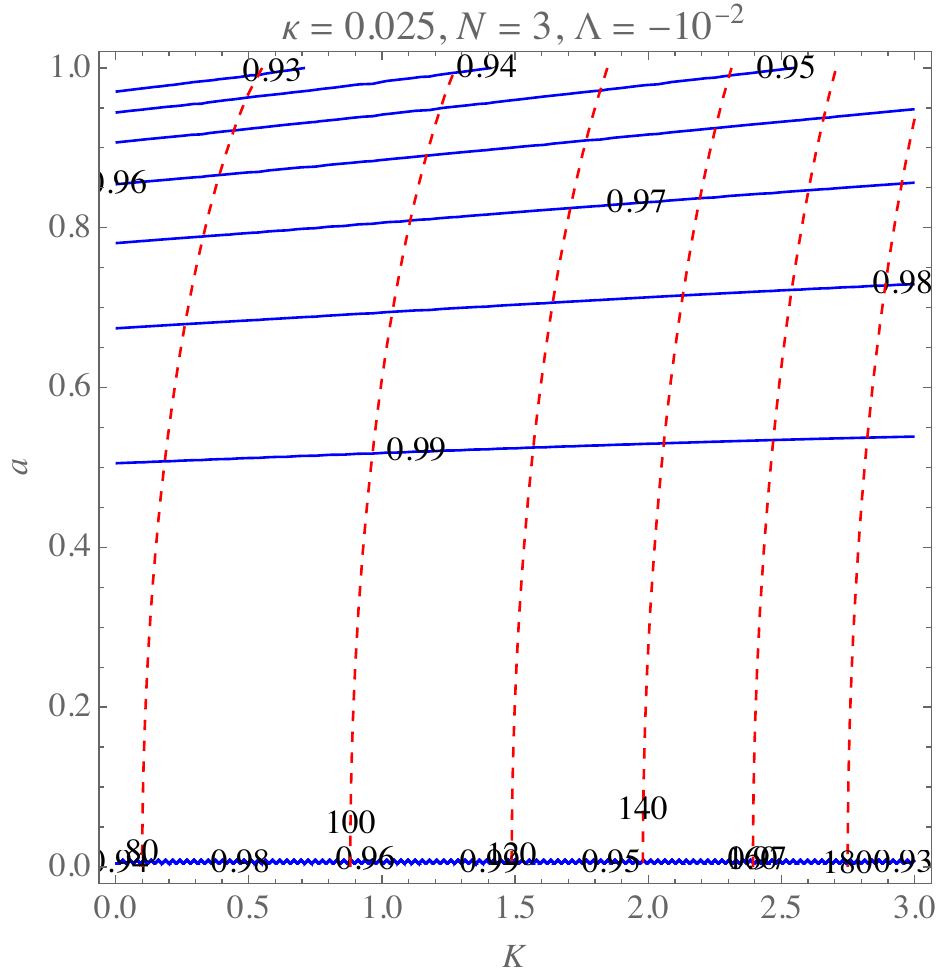} (a)
    \includegraphics[width=7cm]{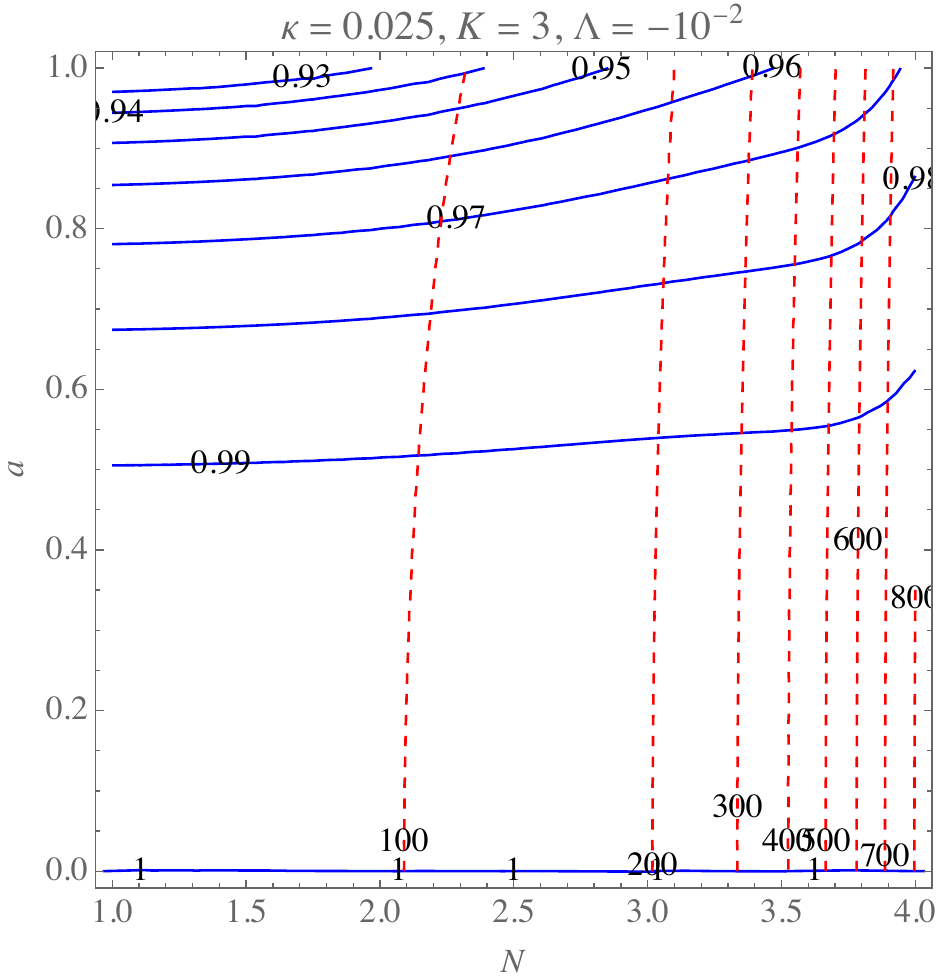} (b)
    \caption{Contour plots of $A_s$ (dashed red curves) and $D_s$ (blue curves), plotted for $\theta_o=\pi/4$ in the parameter spaces (a) $(K,a)$ when $N$ is fixed, and (b) $(N,a)$ when $K$ is fixed. The unit of length has been taken as the BH mass $M$.}
    \label{fig:AsDs}
\end{figure}
%

\subsection{Constraints from the \textit{EHT} observations for M87* and Sgr A*}

The shadow images of two supermassive BHs, M87* and Sgr A*, captured by the EHT collaboration \citep{EventHorizonTelescope:2019dse,EventHorizonTelescope:2019ggy,EventHorizonTelescope:2022wkp,EventHorizonTelescope:2022xqj}, provide compelling motivation for studying BH shadows, which offer significant scientific insights. These observations enable us to estimate the parameters of BHs within the context of various modified or alternative gravity models, helping to identify which models are most consistent with the observed data. In this section, we apply the constraints derived from the EHT observations of M87* and Sgr A* to analyze the parameters of our rotating BH model. Specifically, we utilize the angular diameters of these BHs, as measured by the EHT collaboration, to constrain the model. The angular diameter of a BH's shadow, observed from a distance $d$, can be expressed as \citep{afrin_parameter_2021,zahid_shadow_2023}
\begin{equation}
    \theta_d = 2 \frac{R_a}{d}, \quad R_a = \sqrt{\frac{A}{\pi}},
    \label{eq:thetad}
\end{equation}
where $R_a$ represents the areal radius of the shadow, and $d$ is the observer's distance to the BH. Using Eq. \eqref{Area1}, we find that the angular diameter of the shadow depends on the BH parameters and the observation angle, with an implicit dependence on the BH mass. In our analysis, we treat the supermassive BHs M87* and Sgr A* as RASN-BHs and compare the theoretical shadow predictions with the observational data from the EHT. The mass of M87* and its distance from Earth are taken as $M = 6.5 \times 10^9 M_\odot$ and $d = 16.8 \, \text{Mpc}$, respectively \cite{EventHorizonTelescope:2019pgp,EventHorizonTelescope:2019ggy}. For simplicity, the uncertainties associated with the mass and distance measurements of these supermassive BHs are not considered in the calculations. The angular diameter of the shadow image for M87* is reported as $\theta_d = (42 \pm 3) \, \mu \text{as}$ at the $1\sigma$ confidence level \cite{EventHorizonTelescope:2019dse}. In Fig. \ref{fig:M87ak}, we illustrate the mutual behavior of $a/M$ and $K$ using the theoretical angular diameter in Eq. \eqref{eq:thetad}, for the observational data inferred for M87*, considering two distinct values for the $N$-parameter and two inclinations, $\theta_o = 90^\circ$ and $17^\circ$. Based on the EHT observational results from the shadow images of M87*, it has been shown that the spin parameter of M87* is approximately $a \approx (0.9 \pm 0.05)M$ \cite{Tamburini:2019vrf}, while $\theta_o = 17^\circ$ \cite{Daly:2023axh}. 
\begin{figure}[h]
    \centering
    \includegraphics[width=8cm]{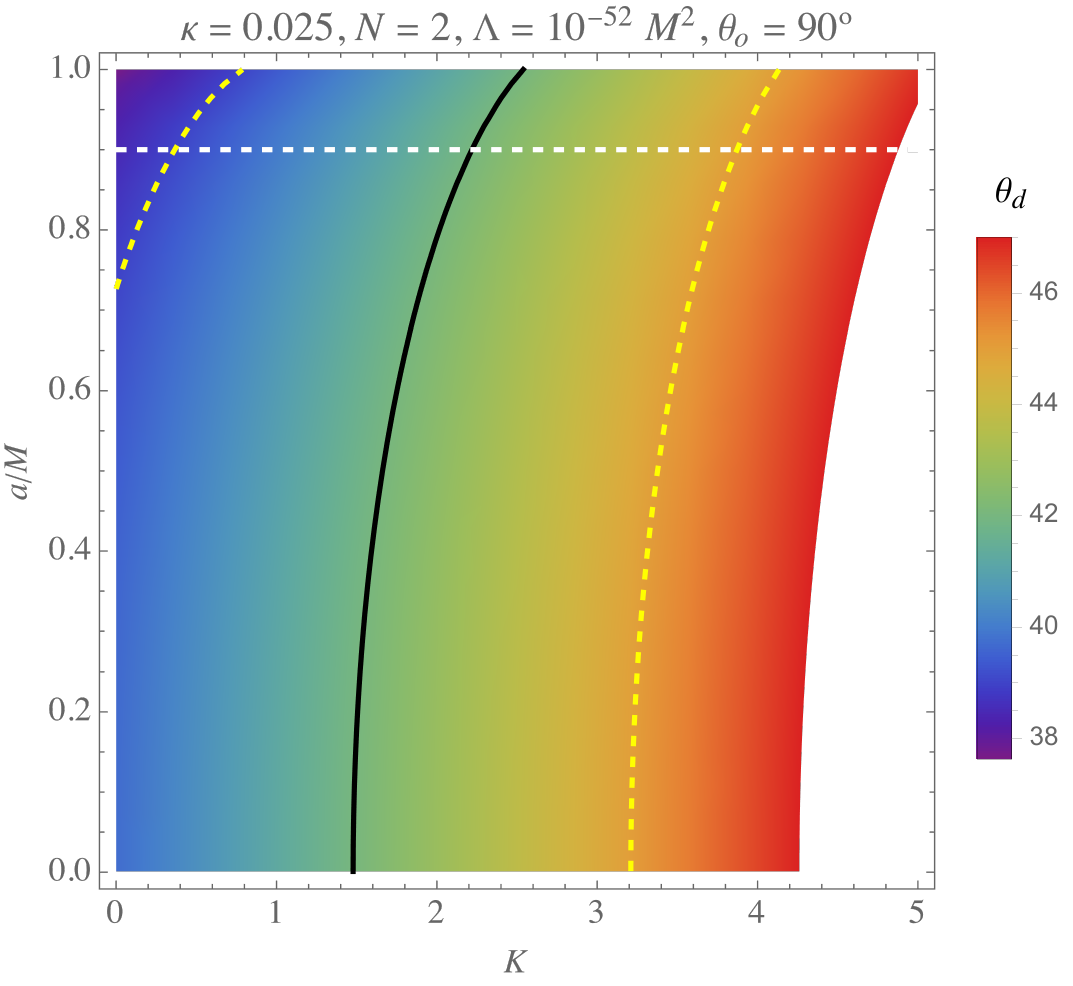} (a)
    \includegraphics[width=8cm]{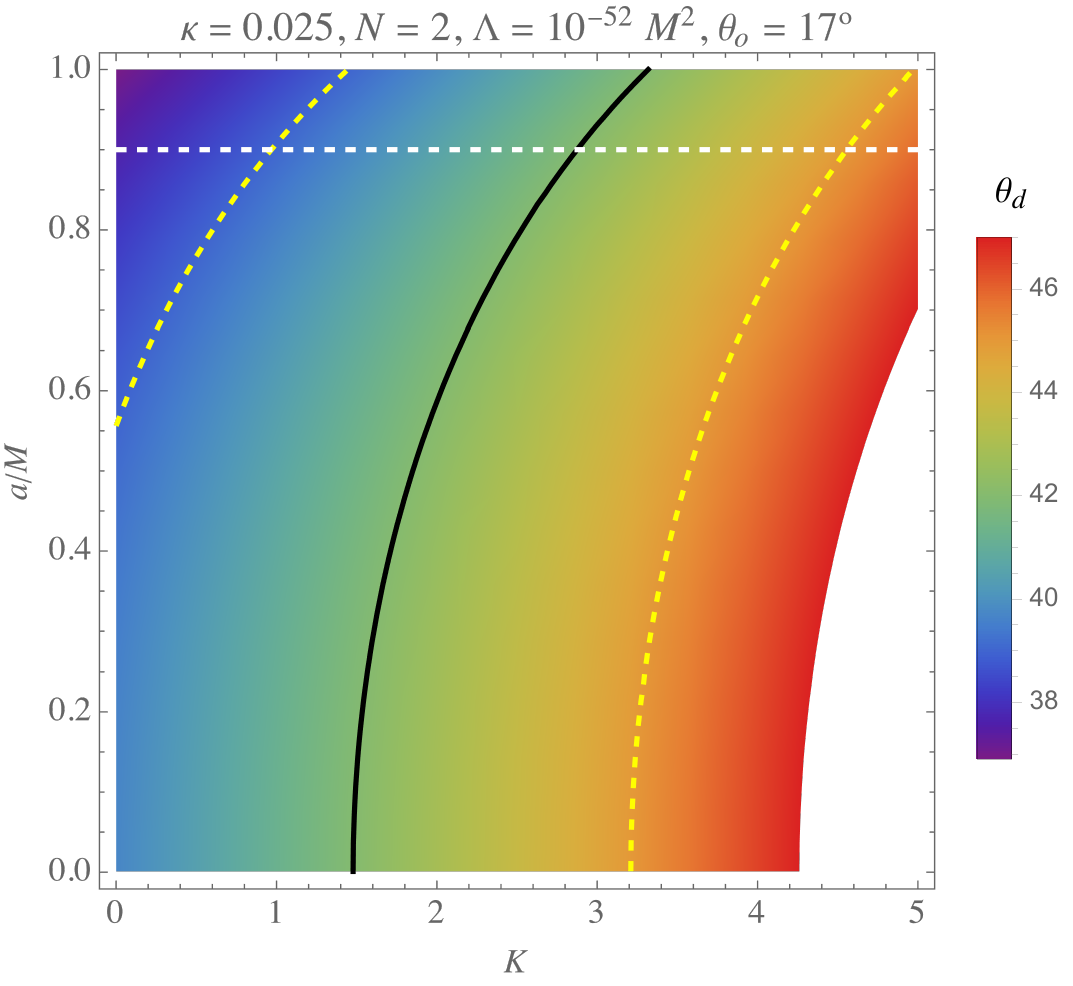} (b)
    \includegraphics[width=8cm]{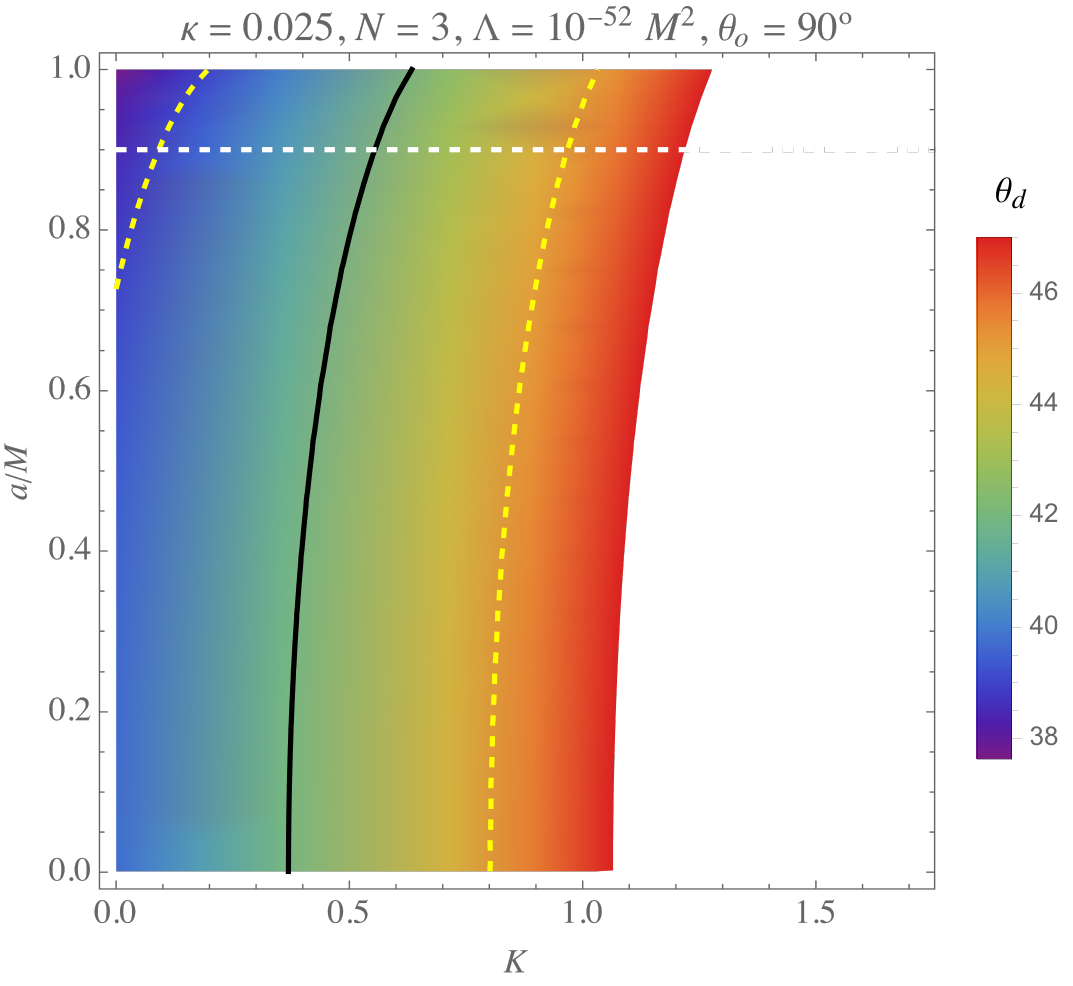} (c)
    \includegraphics[width=8cm]{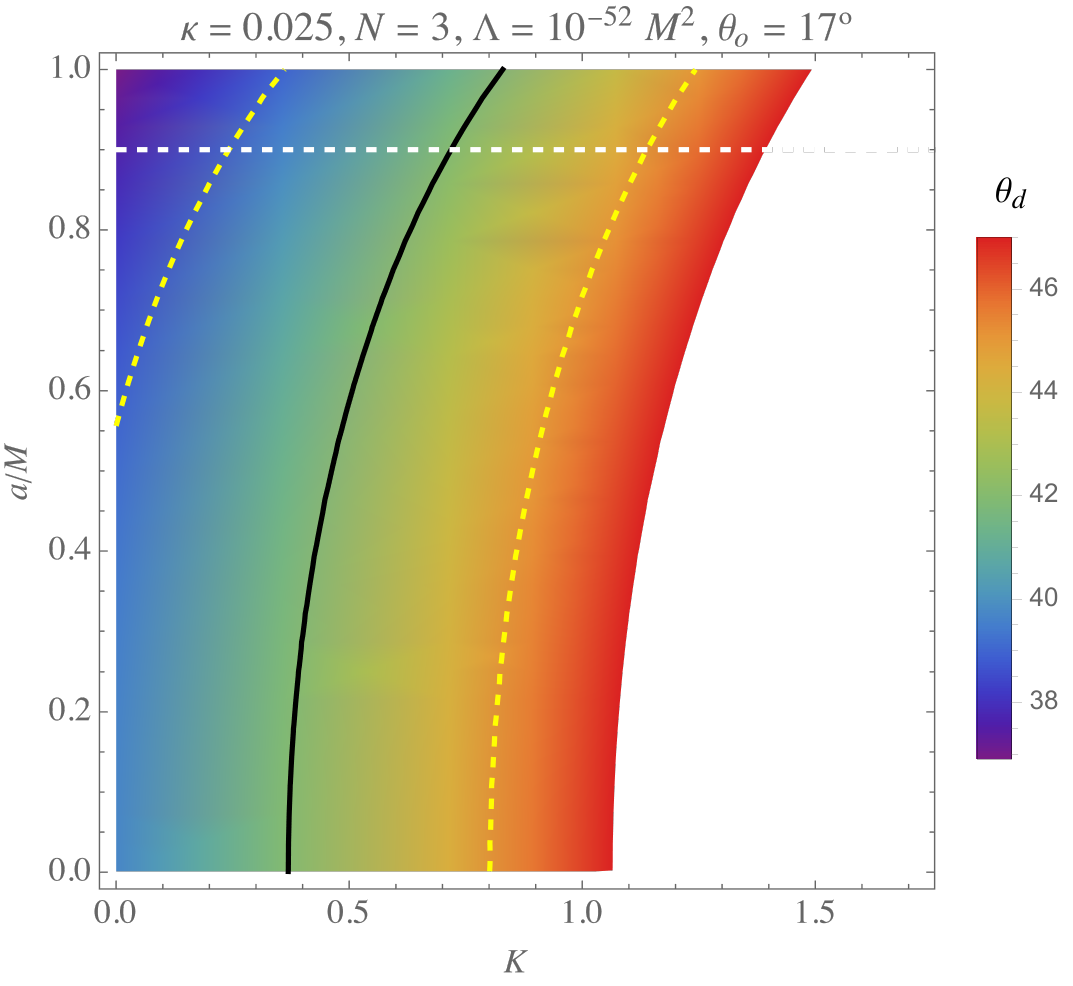} (d)
    \caption{Angular diameter $\theta_d$ for the RASN-BH shadows as a function of parameters $a/M$ and $K$ at inclinations $90^\circ$ (left panels) and $17^\circ$ (right panels), $\kappa=0.025$ and $\Lambda=- 10^{-52}\,\mathrm{m}^{-2}$. The diagrams correspond to (a,b) $N=2$, and (c,d) $N=3$. The black curves describe the borders of the image size at $\theta_d = 42 \, \mu\text{as}$, while the dashed yellow curves correspond to the $1\sigma$ uncertainties $\pm 3 \, \mu\text{as}$, measured for the angular diameter of M87* reported by the EHT. The white dashed line corresponds to $a = 0.9M$.}
    \label{fig:M87ak}
\end{figure}
The constraints on the $K$-parameter depend on the chosen theory, particularly the number of flavors, $N$. Based on the diagrams in Fig. \ref{fig:M87ak}, within the $\mathrm{SU}(2)$ symmetry group and for small values of $\kappa \leq 0.03$, we can constrain $0.967 \leq K \leq 4.556$, while within the $\mathrm{SU}(3)$ symmetry group, the constraint becomes $0.242 \leq K \leq 1.139$. Therefore, we conclude that the reliable range for the $K$-parameter, in light of the observational data from the EHT for M87*, shrinks for models with larger flavor numbers.

Similarly, constraints for the shadow image of Sgr A* can be derived from EHT observations. The angular diameter of the shadow of the supermassive BH Sgr A* is reported as $\theta_o = (48.7 \pm 7) \, \mu\text{as}$ \cite{EventHorizonTelescope:2022wkp}. The mass of Sgr A* and its distance from the solar system are approximately $M = 4 \times 10^6 M_\odot$ and $d = 8 \, \text{kpc}$, respectively \cite{EventHorizonTelescope:2022exc,EventHorizonTelescope:2022xqj}. In Fig. \ref{fig:SgrAak}, the constraints obtained from Sgr A* are illustrated for the two cases under consideration. The fitted density plots of the angular diameter of the BH fall within the measured angular diameter of Sgr A*, $\theta_d = (48.7 \pm 7) \, \mu\text{as}$. Furthermore, recent analysis of Sgr A* suggests that its spin should be $a = (0.9 \pm 0.06)M$ \citep{Daly:2023axh}. 
\begin{figure}[h]
    \centering
    \includegraphics[width=8cm]{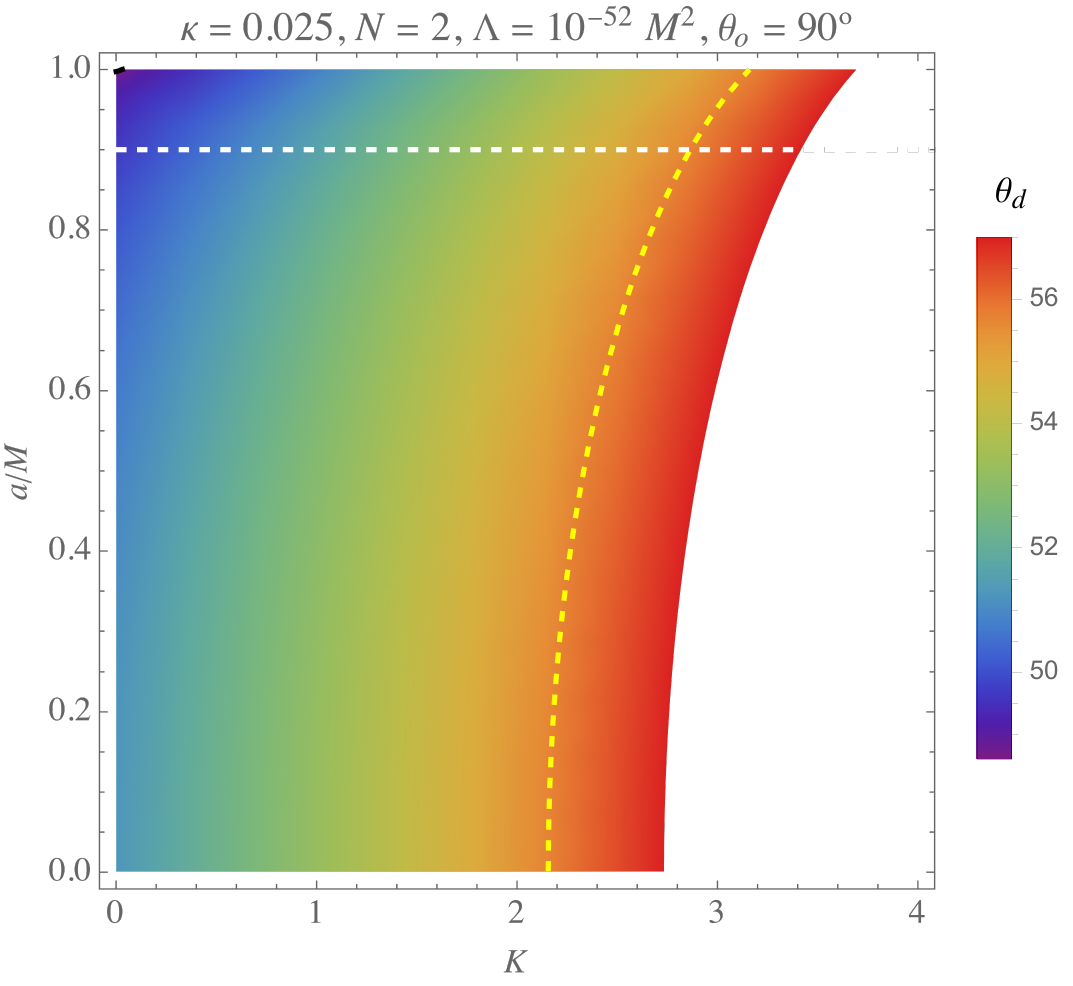} (a)
    \includegraphics[width=8cm]{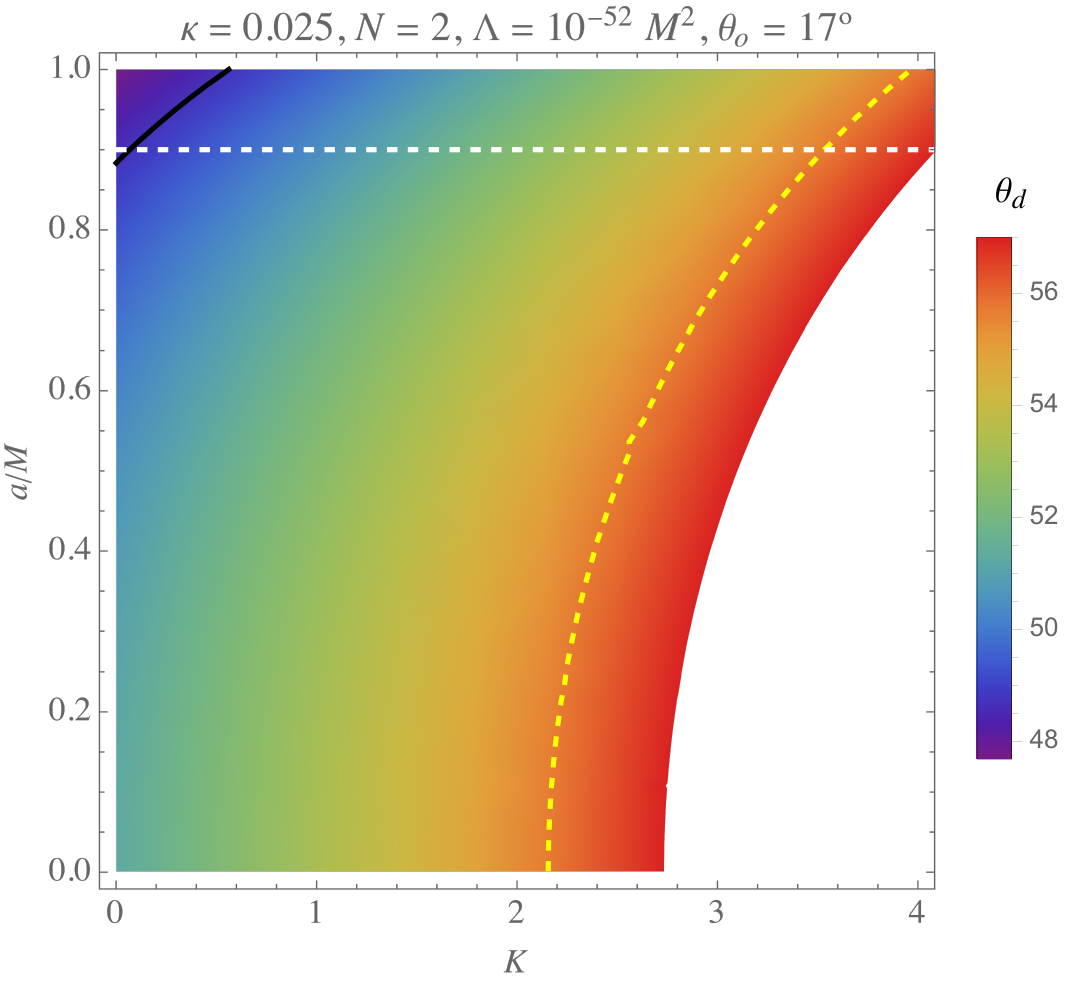} (b)
    \includegraphics[width=8cm]{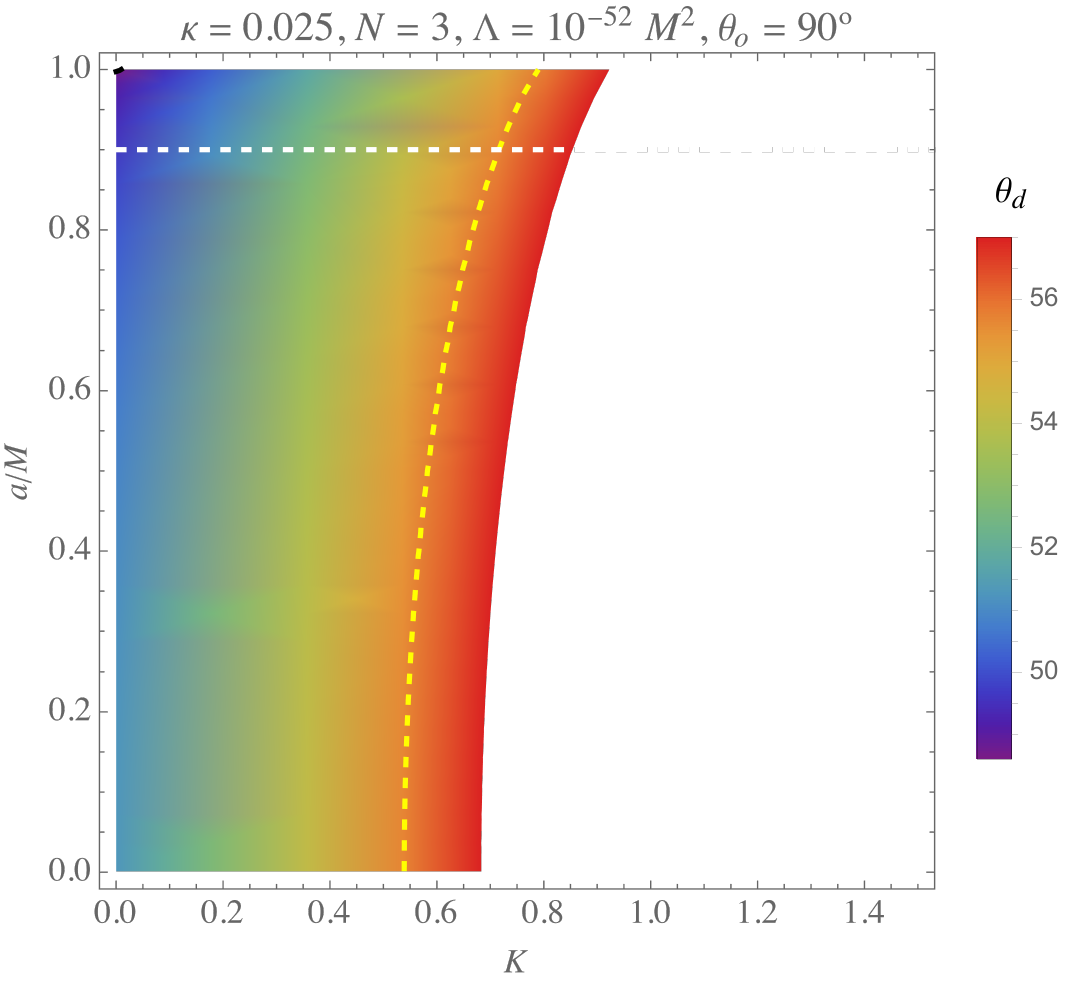} (c)
    \includegraphics[width=8cm]{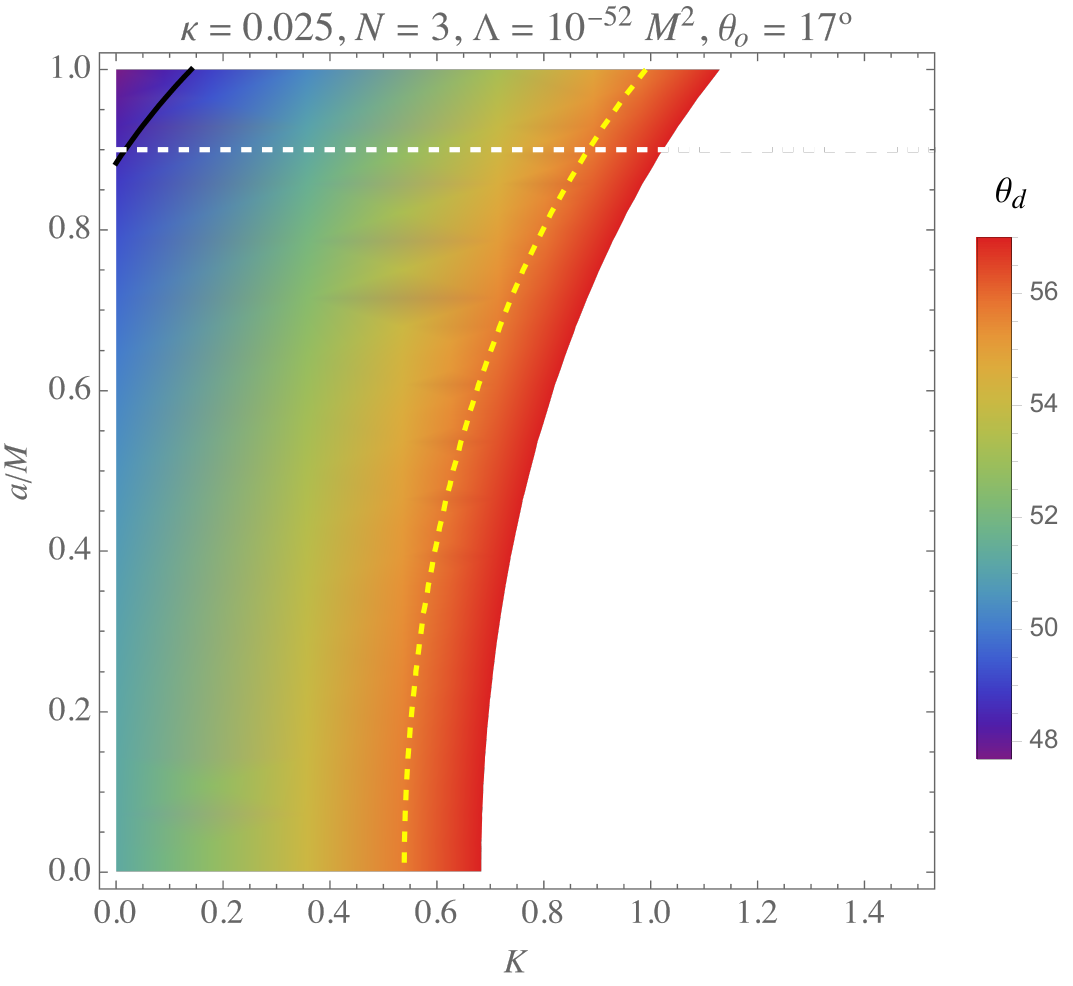} (d)
    \caption{Angular diameter $\theta_d$ for the BH shadows as a function of parameters $a/M$ and $K$ at inclinations $90^\circ$ (left panels) and $17^\circ$ (right panels), $\kappa=0.025$ and $\Lambda=-10^{-52}\,\mathrm{m}^{-2}$. The diagrams correspond to (a,b) $N=2$, and (c,d) $N=3$. The black curve in panels (b,d) describes the border of the image size at $\theta_d = 48.7 \, \mu\text{as}$. The dashed yellow curves correspond to the $+1\sigma$ uncertainty, $+7 \, \mu\text{as}$, measured for the angular diameter of Sgr A* reported by the EHT. The white dashed line corresponds to $a = 0.9 M$.}
    \label{fig:SgrAak}
\end{figure}
Based on these diagrams, one can impose the constraints $0 \leq K \leq 3.534$ for the $\mathrm{SU}(2)$ theory, and $0 \leq K \leq 0.883$ for the $\mathrm{SU}(3)$ theory. Furthermore, as shown in Fig. \ref{fig:EHTaN} for the particular case of $\theta_o = 17^\circ$, the reliable values for the integer within the observational data are $N = 1, 2$, which includes the Kerr-AdS solution.
\begin{figure}[h]
    \centering
    \includegraphics[width=8cm]{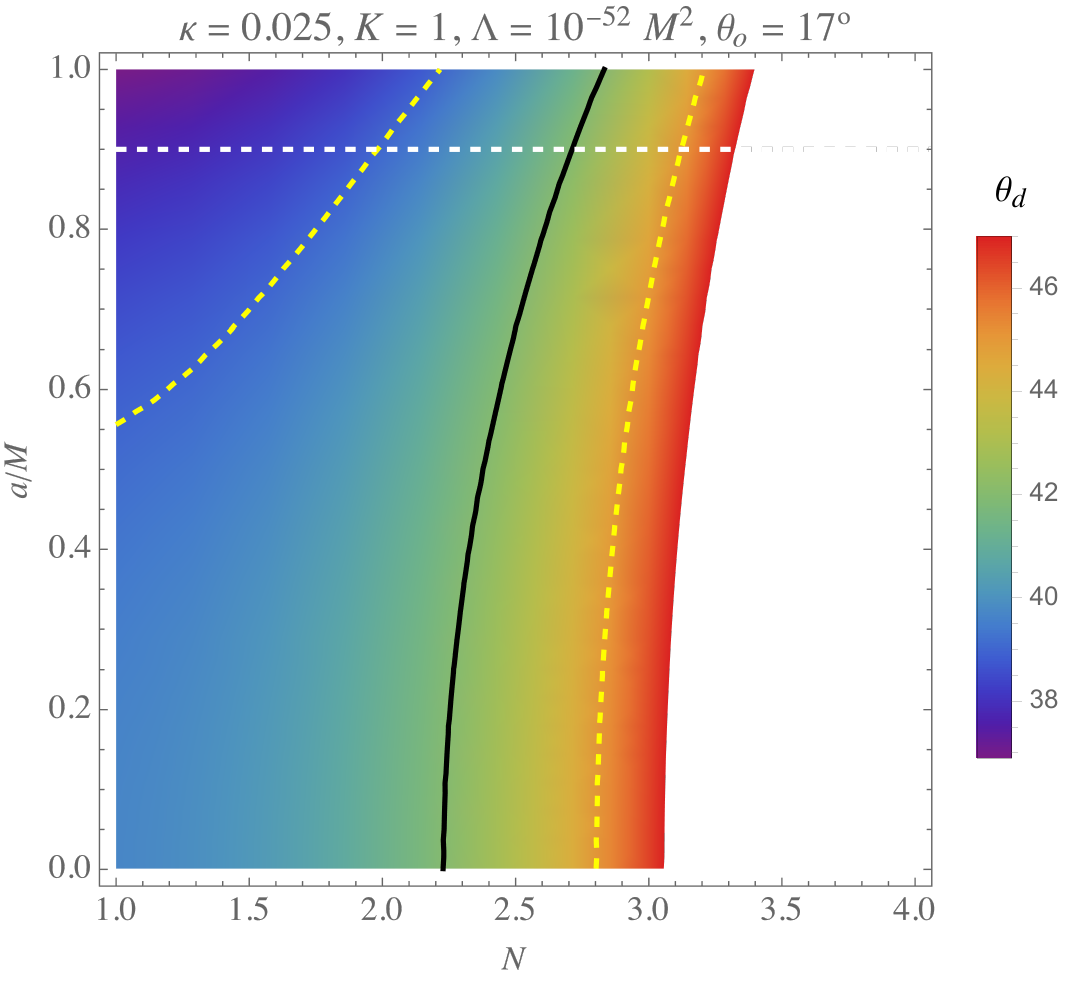} (a)
    \includegraphics[width=8cm]{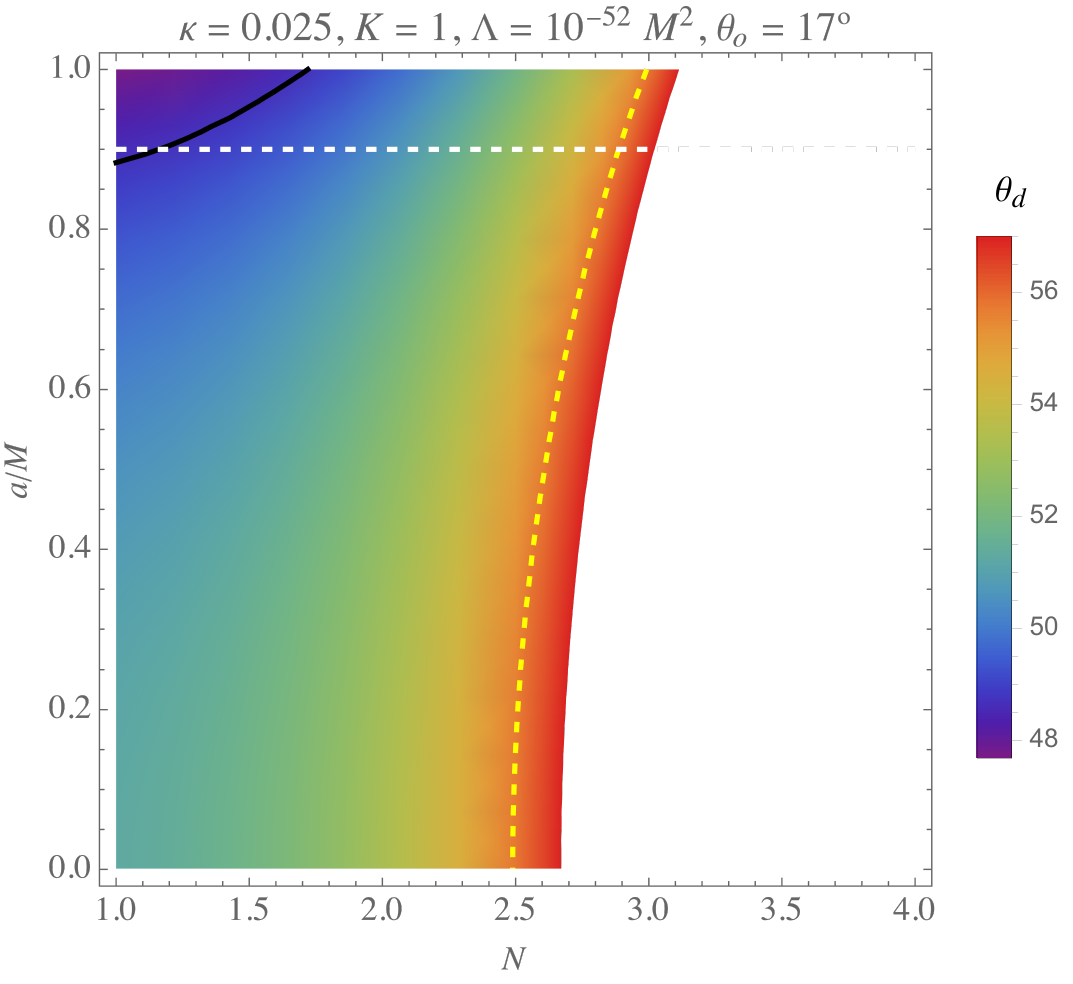} (b)
    \caption{Angular diameter $\theta_d$ for the RASN-BH shadows as a function of parameters $a/M$ and $N$, for $K=1$, at the inclination $17^\circ$, within the data for (a) M87*, and (b) Sgr A*. The color coding is the same as those in Figs. \ref{fig:M87ak} and \ref{fig:SgrAak}, with the constraint for the reliable range of the $K$-parameter for both supermassive BHs indicated in the range of observed $\theta_d$.}
    \label{fig:EHTaN}
\end{figure}
%

\section{The energy emission rate}\label{sec:energyEmission}

From a quantum mechanical standpoint, near a BH's event horizon, particles can be both created and annihilated. Those with positive energy can escape the event horizon through tunneling. This mechanism drives BHs to emit radiation, which could ultimately lead to their evaporation. This phenomenon, known as Hawking radiation, arises due to quantum effects, where BHs radiate thermally, progressively losing mass and energy until they eventually disappear \cite{Hawking:1974rv}. At high energies, Hawking radiation typically emerges within a finite cross-sectional area, denoted by $\sigma_l$. For distant observers far from the BH, this cross-section approaches the shadow cast by the BH \cite{belhaj_deflection_2020, wei_observing_2013}. It has been established that $\sigma_l$ is directly related to the area of the photon ring and can be approximated as \cite{wei_observing_2013, decanini_fine_2011, li_shadow_2020}
\begin{equation}
\sigma_l \approx \pi R_{s}^2.
    \label{eq:sigmal}
\end{equation}
Consequently, the energy emission rate of the BH is given by
\begin{equation}
\Omega \equiv \frac{\mathrm{d}^2 E(\varpi)}{\mathrm{d}\varpi \, \mathrm{d} t} = \frac{2\pi^2 \sigma_l}{e^{\varpi / T_\mathrm{H}^+} - 1} \varpi^3 \approx \frac{2\pi^3 R_{s}^2 \varpi^3}{e^{\varpi / T_\mathrm{H}^+} - 1},
    \label{eq:emissionrate}
\end{equation}
where $\varpi$ represents the emission frequency, and $T_\mathrm{H}^+ = {\tilde{\kappa}_g}/{2\pi}$ is the Hawking temperature at the event horizon, where
\begin{equation}
\tilde{\kappa}_g = \left. \frac{\Delta_r'(r)}{2\left(a^2 + r^2\right)} \right|_{r_+},
    \label{eq:kappa}
\end{equation}
is the surface gravity at the event horizon. It is straightforward to verify that for zero spin parameter (i.e., $a = 0$), this quantity reduces to $\tilde{\kappa}_g = f'(r_+)/2$, which is the surface gravity of the event horizon for static BHs. In Fig. \ref{fig:Omega}, we show examples of the behavior of $\Omega$ as a function of the frequency $\varpi$ for the RASN-BH model.
\begin{figure}[h]
    \centering
    \includegraphics[width=5.4cm]{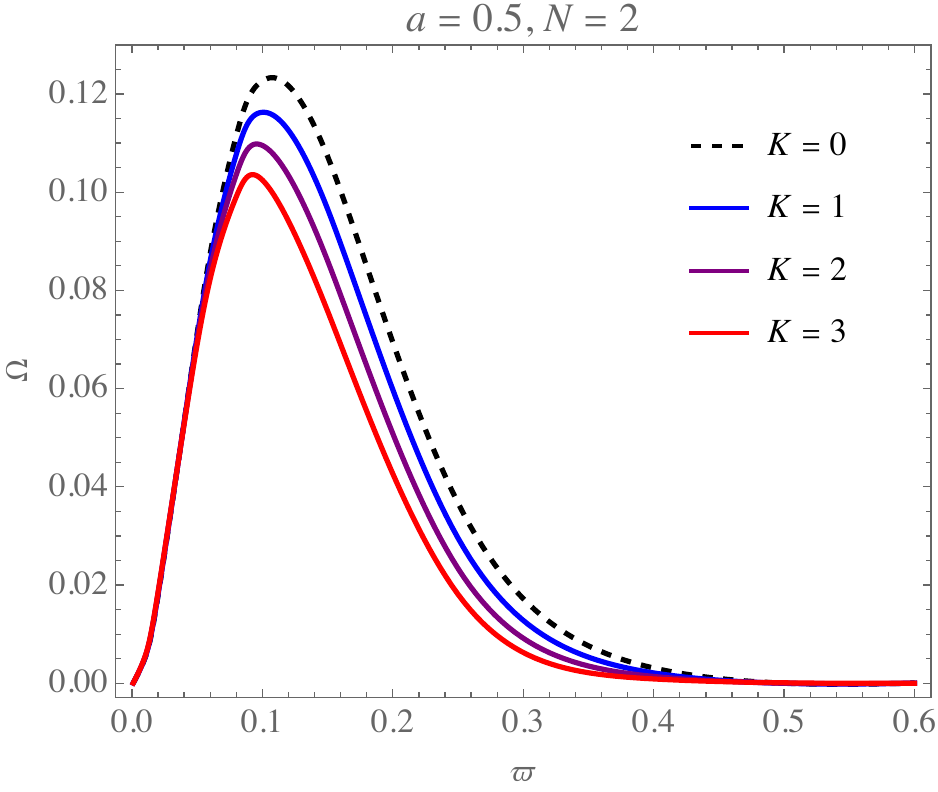} (a)
     \includegraphics[width=5.4cm]{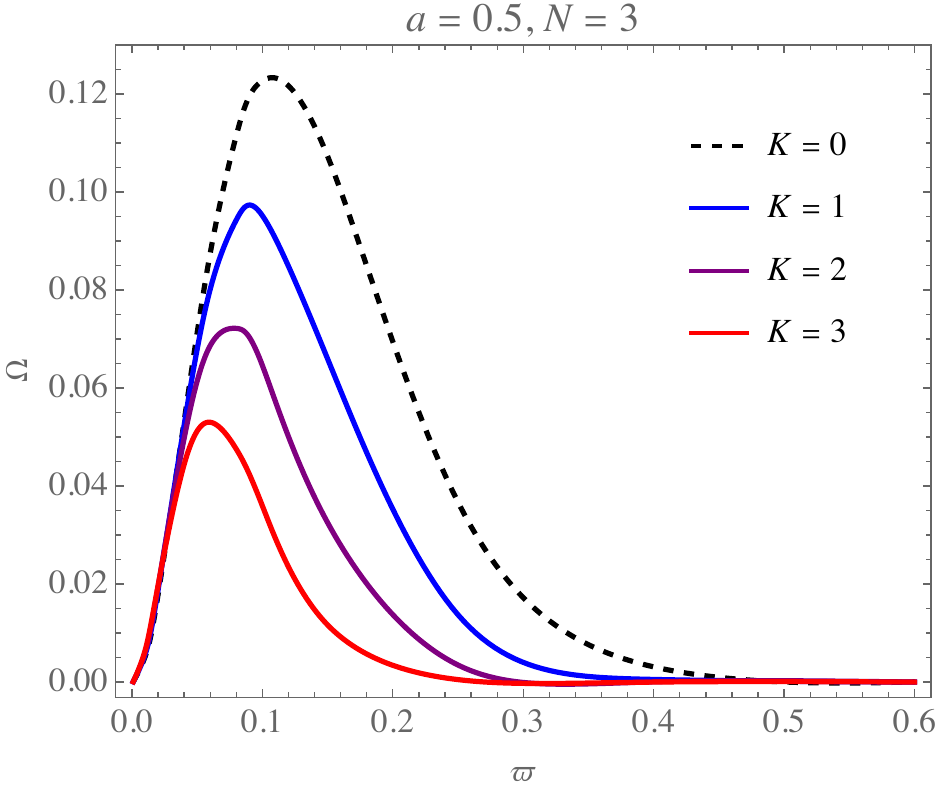} (b)
      \includegraphics[width=5.4cm]{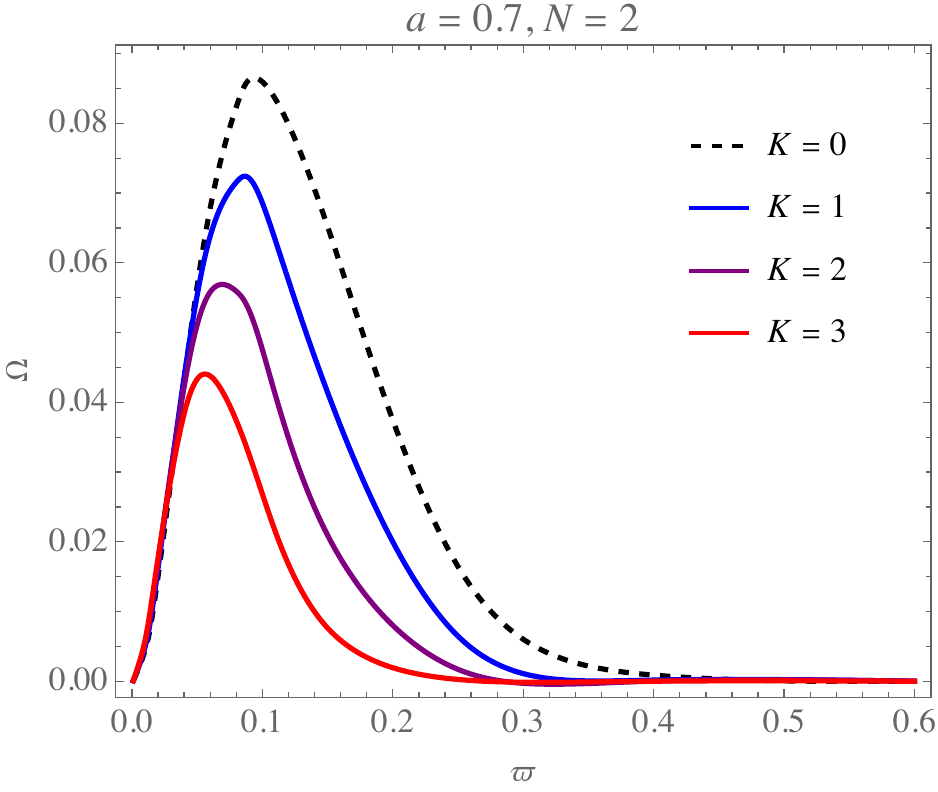} (c)
      \includegraphics[width=5.4cm]{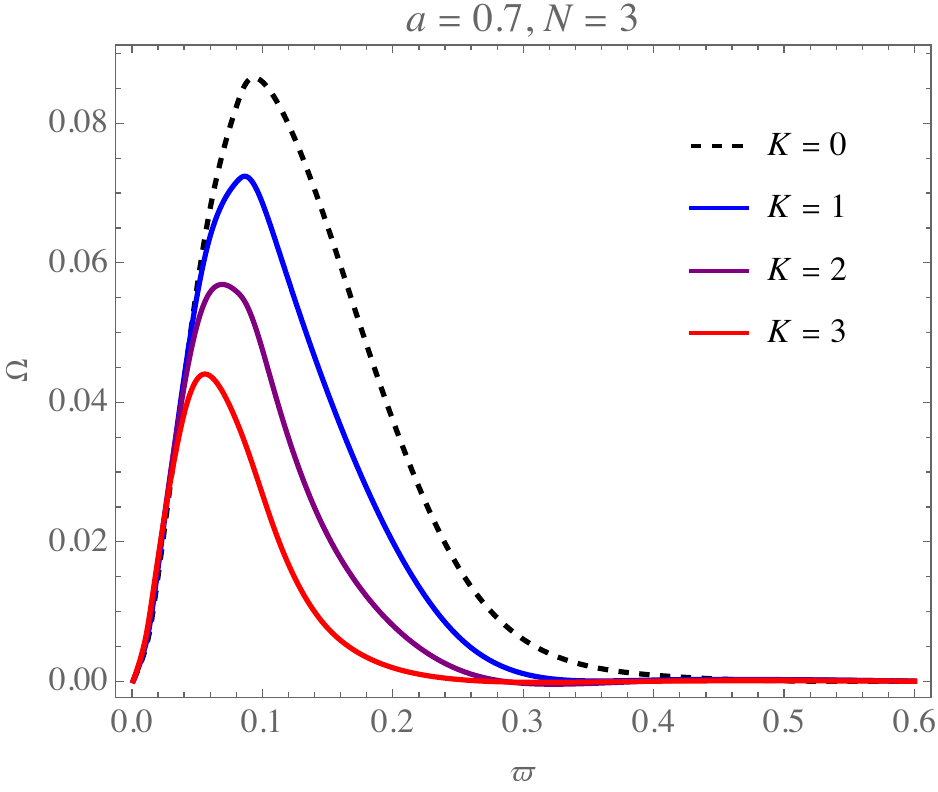} (d)
      \includegraphics[width=5.4cm]{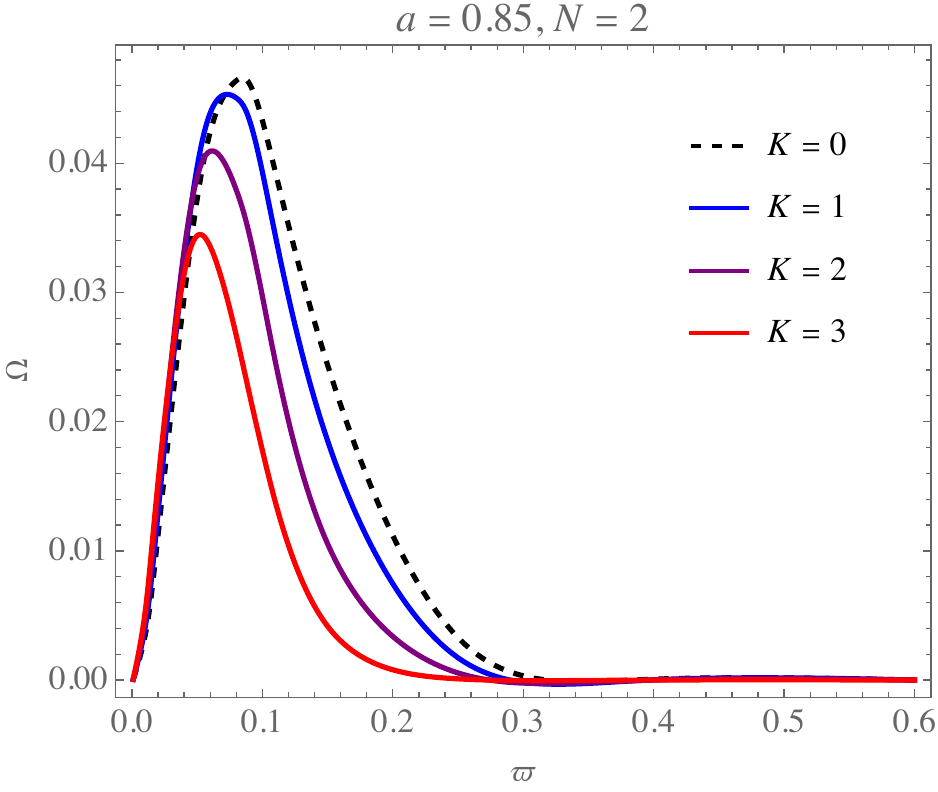} (e)
      \includegraphics[width=5.4cm]{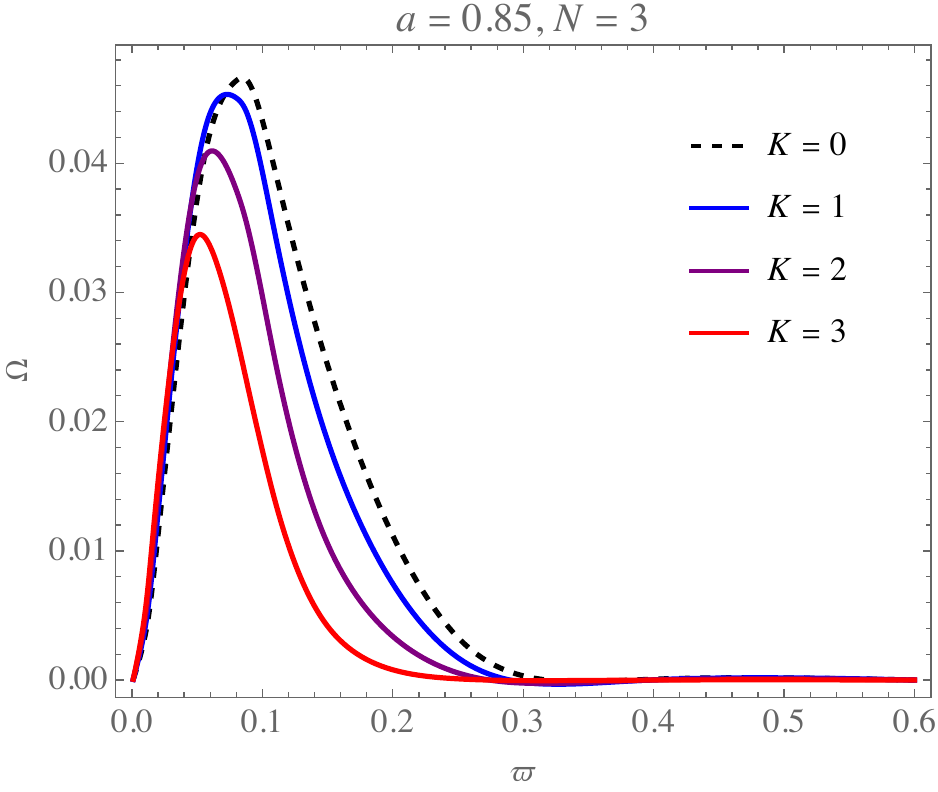} (f)
    \caption{The profiles of the energy emission rate with respect to changes in the frequency $\varpi$, for different cases of the spin and $K$-parameters, assuming $\kappa=0.025$, $\Lambda=-0.01$, and $\theta_d=17^\circ$, for the two relevant cases of $N=2,3$. The units of length are taken as the BH mass $M$.}
    \label{fig:Omega}
\end{figure}
As seen in the diagrams, faster-rotating BHs emit less energy. Furthermore, for all values of $N$, a larger $K$-parameter results in a reduced energy emission rate. For slower-rotating BHs, the Kerr-AdS BH exhibits the highest evaporation rate. However, this behavior evolves as the spin parameter increases (see diagrams (e) and (f)). Specifically, the energy emission rate profile of the RASN-BH for each value of the $K$-parameter approaches that of the Kerr-AdS BH and eventually surpasses it with an increase in the spin parameter. Additionally, it is evident that as $a$ increases, the profiles become more similar, regardless of the value of $N$. {Note also that, as inferred from the diagrams, for a fixed spin parameter, the emission rate decreases with an increase in the flavor number $N$. This suggests that theories with larger $N$ may result in more stable black holes.
}

\section{CONCLUSIONS AND OUTLOOKS}\label{sec:conclusions}

In this work, we have investigated the shadow characteristics of rotating BHs in the Einstein-SU(N)-NLSM within an asymptotically AdS spacetime. Our analysis began with a detailed exploration of the static BH solution, where we examined the causal and asymptotic structures. Using the MNJA, we extended this solution to obtain its rotating counterpart, referred to as the RASN-BH. The properties of this solution, including the horizon structure and ergoregions, were analyzed in depth. 

Utilizing the Lagrangian formalism, we derived the geodesic equations for photons and obtained the BH shadow as observed on the celestial plane. Our results highlight the influence of the BH parameters $K$ and $N$ on the shadow size and shape. By comparing these predictions with the EHT observational data for M87* and Sgr A*, we imposed constraints on the allowed values of $K$ and $N$. Specifically, our analysis shows that larger values of $N$ lead to an increase in the shadow size, while $K$ plays a significant role in the deformation of the shadow. 

The observables associated with the BH shadow, such as the distortion parameter $\delta_s$, shadow area $A_s$, and oblateness $D_s$, were systematically analyzed. We observed that the spin parameter $a$ leads to an asymmetry in the shadow, resulting in a characteristic D-shaped outline. The parameter $K$ was found to modify the shadow size, with increasing $K$ leading to a larger shadow. Furthermore, an increase in the inclination angle $\theta_o$ enhances the deformation of the shadow. 

By applying the EHT constraints, we established a range of allowed values for the model parameters. For M87*, our results suggest that within the SU(2) symmetry group and for $\kappa \leq 0.03$, the constraint on $K$ is $0.967 \leq K \leq 4.556$. For the SU(3) case, the allowed range is narrower, $0.242 \leq K \leq 1.139$. Similarly, for Sgr A*, we found that $0 \leq K \leq 3.534$ for the SU(2) model and $0 \leq K \leq 0.883$ for SU(3), demonstrating the impact of the flavor number $N$ on the BH shadow. 

Additionally, we investigated the energy emission rate of the RASN-BH, which provides insight into its thermodynamic properties. Our analysis indicates that as the spin parameter $a$ increases, the energy emission rate decreases, suggesting that rapidly rotating BHs in this model have lower evaporation rates. The behavior of the energy emission rate also depends on the parameters $K$ and $N$, with higher values leading to reduced radiation. 

Our study, hence, provides a comprehensive analysis of rotating BHs in the Einstein-SU(N)-NLSM, shedding light on the impact of the model parameters on BH shadows and their astrophysical implications. Future work may extend this investigation by considering additional modifications, such as quantum corrections, higher-dimensional extensions, or alternative observational constraints beyond the EHT data. These directions could further refine the viability of the Einstein-SU(N) framework in describing astrophysical BHs and testing deviations from general relativity.

\section*{Acknowledgements}
M.F. is supported by Universidad Central de Chile through project No. PDUCEN20240008. The authors thank Soroush Zare for his discussions and early contributions during the initial stages of this paper's preparation.

\section*{Data Availability}

We have not generated any original data in the due course of this study, nor has any third-party data been analyzed in this article.

\appendix

\section{Notes on the derivation of the rotating spacetime}\label{app:A}

To derive the rotating spacetime \eqref{e1}, we first apply the MNJA to construct the core components of the function \eqref{RASN metric}, and then incorporate the terms associated with the cosmological constant. The technique follows the approach outlined in Refs.~\cite{Azreg:2014, azreg-ainou_static_2014}. 

Let us consider the general static spacetime
\begin{equation}
    \mathrm{d} s^2=-B(r)\mathrm{d} t^2+\frac{\mathrm{d} r^2}{A(r)}+r^2\left(\mathrm{d}\theta^2+\sin^2\theta\mathrm{d}\phi^2\right).
\end{equation}
By assuming the transformation
\begin{equation}
\mathrm{d} t = \frac{\mathrm{d} r}{\sqrt{A(r) B(r)}} + \mathrm{d} u,
\end{equation}
for the advanced null coordinate $u$, the line element can be expressed as
\begin{equation}
\mathrm{d} s^2 = -B(r) \mathrm{d} u^2 - 2\sqrt{\frac{B(r)}{A(r)}} \mathrm{d} r \mathrm{d} u + r^2 (\mathrm{d} \theta^2 + \sin^2\theta \mathrm{d} \phi^2).
\end{equation}
Defining the null tetrad $\bm{e}_{(a)} = \{\bm{l}, \bm{n}, \bm{m}, \bar{\bm{m}}\}$, satisfying the orthogonality conditions, the contravariant components of the metric tensor are given by
\begin{equation}
g^{\mu\nu} = - l^\mu n^\nu - l^\nu n^\mu + m^\mu \bar{m}^\nu + m^\nu \bar{m}^\mu.
\end{equation}
By applying the coordinate transformation
\begin{align}
    r &\to r + \mathrm{i} a \cos\theta,\\
    u &\to u - \mathrm{i} a \cos\theta,
\end{align}
we obtain the null tetrad components in the complex coordinate system, followed by the replacements
\begin{equation}
    A(r) \to \mathcal{A}(r,\theta,a), \quad B(r) \to \mathcal{B}(r,\theta,a), \quad r^2 \to \Psi(r,\theta,a).
\end{equation}
The stationary line element in advanced null coordinates is obtained as
\begin{eqnarray}
\mathrm{d} s^{2} &=& -\mathcal{B} \mathrm{d} u^{2} - 2\sqrt{\frac{\mathcal{B}}{\mathcal{A}}} \mathrm{d} u \mathrm{d} r + \Psi \mathrm{d} \theta^2 - 2 a \sin^2\theta \left(\sqrt{\frac{\mathcal{A}}{\mathcal{B}}} - \mathcal{A} \right) \mathrm{d} u \mathrm{d} \phi \nonumber\\
&&+ 2 a \sin^2\theta \sqrt{\frac{\mathcal{A}}{\mathcal{B}}} \mathrm{d} r \mathrm{d} \phi + \sin^2\theta \left[ \Psi + a^2 \sin^2\theta \left(2\sqrt{\frac{\mathcal{A}}{\mathcal{B}}} - \mathcal{A}\right) \right] \mathrm{d} \phi^2.
\end{eqnarray}
Transforming to Boyer-Lindquist coordinates through
\begin{align}
    \mathrm{d} u &= \mathrm{d} t + \chi_1(r) \mathrm{d} r,\\
    \mathrm{d} \phi &= \mathrm{d} \phi + \chi_2(r) \mathrm{d} r,
\end{align}
and choosing
\begin{align}
    \chi_1(r) &= -\frac{\mathscr{K}(r) + a^2}{\Delta_r(r)},\\
    \chi_2(r) &= -\frac{a}{\Delta_r(r)},
\end{align}
where $\Delta_r(r) = r^2 A(r) + a^2$ and $\mathscr{K}(r) = r^2 \sqrt{A(r)/B(r)}$, we obtain the Kerr-like metric
\begin{multline}
\mathrm{d} s^2 = \frac{\Psi}{\rho^2} \Biggl\{ - \left( \frac{\Delta_r - a^2 \sin^2\theta}{\rho^2} \right) \mathrm{d} t^2 + \frac{\rho^2}{\Delta_r} \mathrm{d} r^2 + \rho^2 \mathrm{d} \theta^2 + 2 a \sin^2\theta \left( \frac{\Delta_r - a^2 - \mathscr{K}}{\rho^2} \right) \mathrm{d} t \mathrm{d} \phi \\
+ \sin^2\theta \left[ \rho^2 - a^2 \sin^2\theta \left( \frac{\Delta_r - a^2 (2 - \sin^2\theta) - 2\mathscr{K}}{\rho^2} \right) \right] \mathrm{d} \phi^2 \Biggr\},
\end{multline}
where $a = J/M$ is the spin parameter.

The function $\Psi$ can be determined by solving the nonlinear equation
\begin{equation}
(\mathscr{K} + a^2 y^2)^2 \left( 3 \partial_r \Psi \partial_{yy} \Psi - 2 \Psi \partial_{ryy} \Psi \right) = 3 a^2 \partial_r \mathscr{K} \Psi^2,
\end{equation}
where $y = \cos\theta$. If $\mathscr{K} = r^2$, a possible solution is $\Psi = r^2 + a^2 y^2$.

The MNJA provides an effective means of generating rotating solutions from static ones while ensuring the separability of the Hamilton-Jacobi equation for null geodesics \cite{junior_spinning_2020}. This method has been extensively used in various gravity theories \cite{Shaikh:2020, kumar_rotating_2018, contreras_black_2019, fathi_spherical_2023}. 

Now based on the static spacetime form given in Eqs. \eqref{metric} and \eqref{e2}, one can infer $\Delta_r=a^2-2Mr+r^2 \left[1-\frac{1}{6} \kappa  K N \left(N^2-1\right)\right]-\frac{1}{3}{\Lambda  r^4}$. However, to fully incorporate the rotating black hole spacetime, including a cosmological constant, we follow the method outlined in Ref.~\citep{xu_kerr-newman-ads_2017}. Assuming the general stationary line element \eqref{e1}, we explicitly derive the components of the field equations for the Einstein-SU(N)-NLSM within this spacetime. This can be efficiently performed using the computation packages such as RGTC~\citep{RGTC} in \textit{Mathematica}\textsuperscript{\tiny{\textregistered}}. By carrying out this procedure, one can verify that the line element \eqref{e1}, together with the expressions in Eqs. \eqref{eq:ingredients}, satisfies the field equations of the theory. Consequently, we conclude that the RASN-BH represents a stationary solution to the model.


\bibliographystyle{ieeetr}
\bibliography{1References}

\end{document}